\documentclass[aps, prx,preprint,notitlepage,superscriptaddress,a4paper,floatfix]{revtex4-2}
\usepackage[english]{babel}
\usepackage{graphicx}
\usepackage{amssymb}
\usepackage{amsmath}
\usepackage{amsfonts}
\usepackage[dvipsnames]{xcolor}
\usepackage[update,prepend]{epstopdf}
\usepackage{bm}
\usepackage{hyperref}
\usepackage{xr}
\begin{document}
\title{Sequence-encoded amphiphilicity controls the interfacial tension of biomolecular condensates}

\author{Nora-Elen Giesinger}
\affiliation{Leibniz-Institut f{\"u}r Polymerforschung Dresden e.V., Hohe Stra{\ss}e 6, 01069 Dresden, Germany}
\affiliation{Institut f{\"u}r Theoretische Physik, Technische Universit{\"a}t Dresden, 01069 Dresden, Germany}

\author{Jens-Uwe Sommer}
\email{sommer@ipfdd.de}
\affiliation{Leibniz-Institut f{\"u}r Polymerforschung Dresden e.V., Hohe Stra{\ss}e 6, 01069 Dresden, Germany}
\affiliation{Institut f{\"u}r Theoretische Physik, Technische Universit{\"a}t Dresden, 01069 Dresden, Germany}
\affiliation{Cluster of Excellence Physics of Life, Technische Universit{\"a}t Dresden, 01062 Dresden, Germany}

\author{Arash Nikoubashman}
\email{anikouba@ipfdd.de}
\affiliation{Leibniz-Institut f{\"u}r Polymerforschung Dresden e.V., Hohe Stra{\ss}e 6, 01069 Dresden, Germany}
\affiliation{Institut f{\"u}r Theoretische Physik, Technische Universit{\"a}t Dresden, 01069 Dresden, Germany}
\affiliation{Cluster of Excellence Physics of Life, Technische Universit{\"a}t Dresden, 01062 Dresden, Germany}

\date{\today}

\begin{abstract}
Biomolecular condensates exhibit surprisingly low surface tensions, despite forming dense, liquid-like phases. We propose that the low surface tension partially originates from the segmental amphiphilicity of intrinsically disordered proteins (IDPs) that results in a surfactant-like behavior. We investigate this mechanism using a minimal coarse-grained IDP model containing only hydrophobic and polar residues, allowing us to systematically vary the polar fraction and, for selected cases, to probe the influence of sequence patterning of polar residues. To control for differences in thermodynamic phase behavior, we determine phase coexistence (binodals) for each sequence and compare systems at matched distances from the critical point. We find a disproportionately high fraction of polar residues at the condensate surface, which is closely correlated with the reduction in surface tension. For the sequences studied here, overall composition alone sets the phase behavior and establishes the overall trend in surface tension: increasing the polar fraction lowers the surface tension, with a polar fraction of one third reducing the surface tension by roughly twofold relative to a purely hydrophobic chain. In addition, sequence patterning enhances interfacial localization of polar residues and can lower surface tension beyond what composition alone achieves, revealing a sequence-dependent contribution to the surface tension that leaves the bulk phase behavior essentially unchanged.
\end{abstract}

\maketitle
\section{Introduction}
Biomolecular condensates are membraneless compartments that organize proteins and nucleic acids spatially and temporally inside cells. They are thought to form through liquid-liquid phase separation (LLPS), in which biomolecules demix into condensed droplets that coexist in equilibrium with a dilute phase. The liquid-like nature of condensates allows them to assemble and dissolve on demand \cite{brangwynne:sci:2009, hyman:arc:2014, wang:bpr:2025, banani:nrmcb:2017} to regulate local concentrations, thereby controlling cellular processes such as stress response, signal transduction, and gene regulation in space and time \cite{franzmann:csp:2019, su:sci:2016, hnisz:cell:2017, sabari:sci:2018}. Because condensates lack a confining membrane, the interface between biomolecular condensates and their surrounding aqueous phase plays a central role \cite{wang:bpr:2025, folkmann:sci:2021, boeddeker:np:2022, farag:nc:2022, bauer:nc:2024}. The interface contributes to the free-energy cost of forming a condensate, governs how condensates coarsen, fuse, and ripen, and mediates their wetting of membranes or other cellular structures.

The macroscopic physical quantity that characterizes these interfacial properties is the surface tension, which quantifies the free energy needed to create an interface between a condensate and its surrounding phase. Biomolecular condensates typically have extremely low surface tensions, ranging approximately from $10^{-3}$ to $1\,\text{mN}/\text{m}$ \cite{elbaum-garfinkle:pnas:2015, fisher:nc:2020, ghosh:nc:2021, wang:bpr:2021, holland:sm:2023, devarajan:nc:2024, wang:bpr:2025}, and can therefore be readily deformed by cellular forces. The magnitude of the surface tension is set by the molecular interactions that make the interface energetically costly, weighed against thermal energy. This balance is captured by how far the system is from its critical point -- the temperature (or interaction strength) beyond which the condensed and surrounding dilute phases can no longer co-exist and merge into a single homogeneous phase. As the critical point is approached, the two phases become increasingly similar and the surface tension vanishes. Proximity to criticality has therefore been proposed as one reason for the low surface tensions of biomolecular condensates \cite{pyo:pnas:2023}.

However, it is challenging to quantify the role of criticality experimentally, because this analysis requires precise coexistence density and surface tension measurements, ideally near the critical point. Yet such data are available only for relatively few biomolecular condensates \cite{brady:pnas:2017, schuster:nc:2018, wang:bpr:2021, holland:sm:2023}. Further, the reported surface tensions can also depend substantially on the specific measurement method; for example, published measurements for LAF-1 RGG-RGG show a spread of about $75\,\%$ (see Sec.~S01 in the SI) \cite{wang:bpr:2021}. Moreover, the large concentration differences between the dilute and condensed phases typically observed in biomolecular condensates \cite{murthy:nsmb:2019, martin:sci:2020, holland:sm:2023, mccall:nc:2025} indicate that experimentally accessible states are not necessarily asymptotically close to criticality. This is important because the surface tension depends strongly on the phase-density difference, $\Delta \rho$, so that uncertainties regarding the proximity to criticality are significant for the interpretation of $\gamma$. These considerations lead us to the question whether the low surface tensions of biomolecular condensates can be explained solely by their proximity to criticality, or whether their molecular composition provides an additional contribution.

The molecular interactions within biomolecular condensates depend on many coupled features, including protein sequence and valency, composition stoichiometry, charge distributions, salt concentrations, and pH \cite{wang:cell:2018, martin:sci:2020, pyo:iscience:2022, das:pnas:2013}. Because these contributions are intertwined, separating the contribution of any single mechanism to the surface tension is difficult. One of these features, however, is a property of the individual chains alone: the different cohesive and solvation preferences of the constituent residues of intrinsically disordered proteins (IDPs). These proteins, which often drive biomolecular condensation, can be viewed as flexible heteropolymers without structured domains \cite{alston:jpcb:2023, witzky:jcp:2025, witzky:jcp:2025B}. Unlike small-molecule surfactants, however, the hydrophobic and polar residues within IDPs are generally not segregated into well-defined blocks but distributed along the sequence. This distributed patterning can support macroscopic phase separation instead of finite sized clusters (micelles)\cite{israelachvili:ap:2011}. Nevertheless, individual chain segments may still rearrange locally at the condensate surface to reduce the interfacial free energy \cite{golani:cp:2025}. We refer to this local reorganization of amphiphilic segments as ``self-surfactancy'', and propose it as a microscopic mechanism that can contribute to the remarkably low surface tension of biomolecular condensates. 

Testing the idea of self-surfactancy in natural IDPs is difficult, because composition, sequence patterns, and distance from criticality are entangled in a vast sequence space. We therefore turn to a minimal heteropolymer model with only hydrophobic (H) and polar (P) residue types -- a coarse-graining framework that has been shown to qualitatively capture the experimental phase behavior of IDPs \cite{rekhi:jpcb:2023}. Reducing the alphabet to two letters lets us vary the polar fraction (composition) and its arrangement (sequence), while directly measuring phase behavior, surface tension, and interfacial structure. For each sequence, we determine its coexistence curve (binodal) and compare systems at matched thermodynamic states along this curve, thereby controlling the proximity to criticality and exposing how composition and sequence influence self-surfactancy and interfacial tension.

Within this model, we find that \textit{composition} -- the average polar fraction -- sets the phase behavior and the overall trend in surface tension: at matched distance from the critical point, increasing the polar content lowers the surface tension because more polar residues are promoted to the interface. For the systems studied here, the binodals are well described by an effective Flory-Huggins model, where the interaction strength depends mainly on composition. By contrast, the tested sequence variations leave the binodals essentially unchanged, yet strongly affect the surface tension through the extent of polar enrichment at the interface. As a result, a sequence with fewer polar residues can have a lower surface tension than a more polar sequence at the same distance from criticality. Sequence arrangement therefore acts as a largely independent control parameter for the surface tension of biomolecular condensates, tuning the interface without substantially altering bulk phase behavior - a decoupling with consequences for both natural condensates and the design of synthetic ones.

\section{Results}
\label{Sec.Results}


\subsection{Model System}
\label{Sec.Results.Model}

We study a minimal bead-spring heteropolymer model, composed of a linear sequence of H (hydrophobic) and P (polar, hydrophilic) residues. We model the residues as spherical beads, connected through finite extensible springs, where the H and P beads have multivalent pair-interactions. The solvent is modeled implicitly, and the effective interactions between hydrophobic residues (H-H) are modeled though an attractive short ranged pair interaction, while interactions between polar residues (P-P) are excluded volume interactions only. The pair interactions between hydrophobic and polar residues (H-P) are weakly attractive to homogenize the dense phase \cite{rekhi:jpcb:2023}. Further details on the simulation model are provided in the Methods Section.

We first vary the fraction of polar residues in the sequence, $f_\text{P}$, for regular sequence patterns. Specifically, we consider repeating sequences in which $n$ hydrophobic residues are followed by one polar residue, yielding repeating $\text{H}_n\text{P}$ patterns which fit with an integer number in a chain with a fixed length of $M=60$ residues (Fig.~\ref{fig:phase.behavior}a). Here, we fixed $M$ to exclude chain-length effects on the phase behavior and surface tension \cite{schultz:jacs:1952, legrand:jcis:1969, silmore:mp:2017}. We study three regular sequences with increasing $f_\text{P}$: $[\text{H}_{9}\text{P}]_6$ ($f_\text{P}=1/10$), $[\text{H}_{4}\text{P}]_{12}$ ($f_\text{P}=1/5$) and $[\text{H}_{2}\text{P}]_{20}$ ($f_\text{P}=1/3$), together with the purely hydrophobic reference system $[\text{H}_{60}]_1$ ($f_\text{P}=0$). These sequences provide a controlled reference for assessing composition-dependent trends. Uniformly distributing polar residues along the chain suppresses microphase separation, while a sufficient hydrophobic content is required to drive condensation \cite{rekhi:jpcb:2023}.  

To assess the influence of sequence patterning at fixed composition ($f_\text{P} = 1/5$), we compare the regular sequence $[\text{H}_{4}\text{P}]_{12}$ with a more blocky sequence $[\text{H}_{8}\text{P}_{2}]_6$. To probe sequence irregularities that are common in natural IDPs, we additionally consider two irregular sequences, $\text{Seq}_1$ and $\text{Seq}_2$.

To facilitate comparison across different systems, we report all results in reduced units: energies in units of the H-H interaction strength $\varepsilon$, lengths in units of the bead diameter $\sigma$, and surface tensions in units of $\varepsilon/\sigma^2$. Using a mapping of $\sigma=1\,\text{nm}$ and $\varepsilon = k_\text{B}T$ at $T=25^\circ\,\text{C}$, one reduced unit of surface tension ($\varepsilon/\sigma^2$) corresponds to approximately $4.1\,\text{mN}/\text{m}$.


\subsection{Phase Behavior}
\label{Sec.Results.PhaseBehavior}

\begin{figure*}[ht!]
    \centering
    \includegraphics[width=\textwidth]{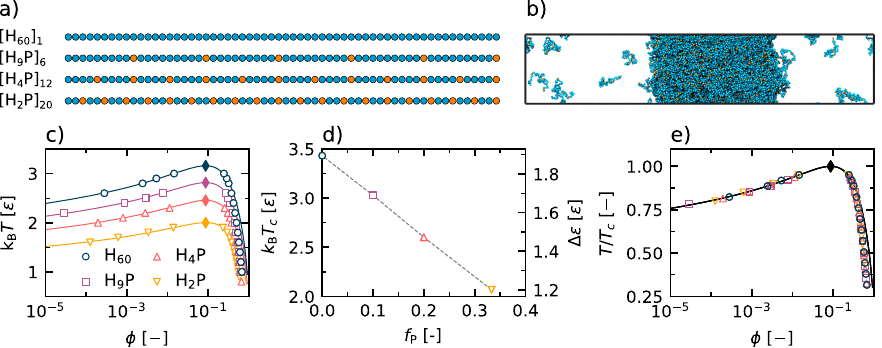}
    \caption{Phase behavior of regular sequences of type $\text{H}_n\text{P}$.
    a) Visual representation of the four sequences, where hydrophobic residues are shown in blue, and polar residues shown in orange.
    b) Simulation snapshot of a slab geometry of the $[\text{H}_9\text{P}]_{6}$ system at $\tau = 0.09$, with a populated dense and dilute phase.
    c) The binodals of the four systems in absolute temperature. The statistical uncertainties of the data are smaller than the symbol size. Solid lines are fits of the standard FH model to our simulation data, and filled diamond symbols mark the critical temperatures obtained from the FH fit.
    d) Critical temperature (left axis), and equivalently the effective site-to-site interaction energy $\Delta \varepsilon$ (right axis), from the FH fits as a function of the polar fraction $f_\text{P}$ in the system. The dashed gray line is a guide to eye only.
    e) Collapse of the binodals after rescaling by their respective critical temperatures, $T_\text{c}$.}
    \label{fig:phase.behavior}
\end{figure*}

Our goal is to quantify how polar residues affect the interfacial free energy of biomolecular condensates.  In cell-like environments, condensates formed by different proteins coexist at the same physiological temperature $T$. However, surface tensions measured at fixed absolute temperature reflect not only sequence- and composition-dependent interactions, but also differences in their proximity to the critical temperature, $T_\text{c}$, as discussed in Ref.~\citenum{pyo:pnas:2023}. To isolate the effect of self-surfactancy, we compare systems at matched thermodynamic conditions, which in practice means comparing the systems at the same proximity to the critical point. We quantify this distance using the conventional reduced temperature
\begin{equation}
    \tau = 1 - T / T_\text{c}\,\,.
    \label{eq:tau_Def}
\end{equation}
We first determine the binodals of the different sequences using direct coexistence simulations in a slab geometry (Fig.~\ref{fig:phase.behavior}b,c).

The phase behavior of the purely hydrophobic reference system, $[\text{H}_{60}]_1$, can be well described by the classical Flory-Huggins (FH) theory \cite{flory:cornell:1953}. For the amphiphilic heteropolymers, however, there are two additional interaction parameters for the H-P and P-P pairs, and a one-component FH treatment must absorb them into one effective parameter. We therefore test whether the coexisting densities of heteropolymer systems can be captured by such a treatment, and how the effective interaction parameter depends on the polar fraction $f_\text{P}$ (composition).

We fit the simulated coexisting densities of each system to the classical FH equation (eq.~\ref{eq.FH}) using the self-consistent solution developed in Ref.~\citenum{qian:acs:2022}. The model contains two fit parameters: an effective chain length, $M_\text{eff}$, and an effective site-to-site interaction energy, $\Delta\varepsilon$, related to the FH interaction parameter via $\chi=\Delta\varepsilon/k_\mathrm{B}T$.  For heteropolymers, $\Delta \varepsilon$ quantifies the effective interaction strength between coarse-grained residues. By optimizing the fit to the $[\text{H}_{60}]_1$ reference system, we found $M_\text{eff}\simeq110$, which we fixed for all sequences. $M_\text{eff}$ is about two times larger than the number of residues, reflecting the off-lattice and implicit solvent nature of our simulation model. Fixing $M_\text{eff}$ leaves a single free parameter for each system, $\Delta\varepsilon$, so that differences in phase behavior are captured entirely through changes in the effective interaction strength. 

As expected, FH theory matches the coexisting densities of the $[\text{H}_{60}]_1$ homopolymer system (Fig.~\ref{fig:phase.behavior}c). The fitted interaction strength for the purely hydrophobic chain is $\Delta\varepsilon_{\mathrm{H}_{60}}=(1.895\pm0.003)\varepsilon$. Remarkably, the same FH homopolymer model also describes the phase behavior of the heteropolymers with excellent accuracy, implying that the amphiphiles behave as homopolymers with weakened effective site-to-site interaction energies $\Delta\varepsilon$. Over the composition-range studied here, $\Delta\varepsilon$ decreases linearly with increasing polar fraction $f_\mathrm{P}$ (Fig.~\ref{fig:phase.behavior}d). This trend is consistent with the mean-field nature of FH theory, in which the distinct H-H, H-P, and P-P interactions are represented by a composition-dependent average interaction parameter. With increasing polar fraction, hydrophobic contacts become less frequent, weakening the effective attraction and thereby lowering $T_\text{c}$.

The employed FH model contains a single energetic scale, $\Delta\varepsilon$, so its coexistence curve obeys van der Waals' law of corresponding states. Thus, the binodals collapse onto one master curve once the temperature is expressed in units of $1/T_\text{c}$, which allows to compare coexisting densities of different systems at matched distance from the critical point (Fig.~\ref{fig:phase.behavior}e). We find good agreement between the dilute branch of the binodals, while minor deviations remain in the dense branches, where systems with higher hydrophobic content exhibit slightly higher volume fractions. Because the binodals collapse so closely onto each other, comparing systems at the same reduced temperature is equivalent to comparing them at the same saturation concentration. The distance from the respective critical temperature, $\tau$, thus serves as a robust descriptor of the thermodynamic state, providing a well-defined basis for comparing the different systems.

Although the FH model describes the binodals over most of the coexistence region, it does not capture the non-mean-field critical behavior near $T_\text{c}$. Since our comparison relies on the reduced temperature $\tau$, we obtain a separate measure of the critical temperature, $T_\text{c,I}$, from the critical scaling of the coexisting-density difference
\begin{equation}
    \Delta\rho = \rho_\text{l} - \rho_\text{v}
    = A\tau^\beta,
\end{equation}
where $\beta \approx 0.326$ is the critical exponent of the three-dimensional Ising universality class. The resulting values of $T_\text{c,I}$ agree with the critical temperatures $T_\text{c}$ obtained from the FH fits to within 3.5\% of its value (see Tables~S1 and S2 in the SI).


\subsection{Surface tension}
\label{Sec.Results.SurfaceTension}

\begin{figure}[ht]
    \centering
    \includegraphics[width=7.5cm]{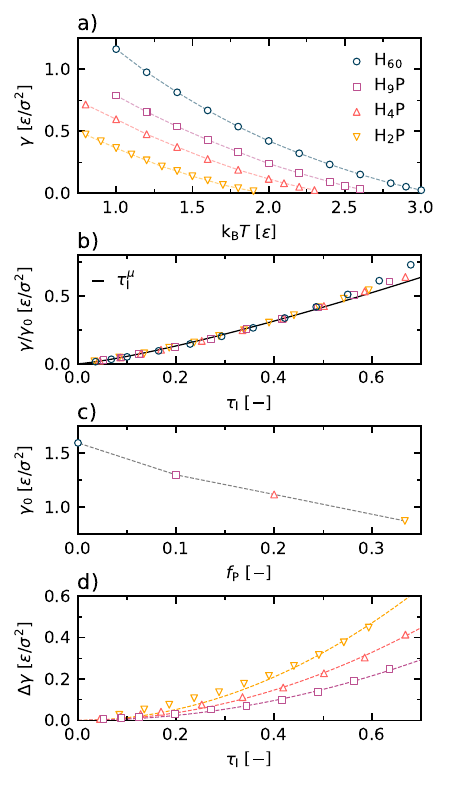}
    \caption{Surface tension of regular sequences of type $\text{H}_{n}\text{P}$.
    a) Surface tension, $\gamma$, as a function of the absolute temperature, $k_\text{B}T$. Open symbols show $\gamma$ from simulations (error bars are smaller than the symbol size), and dotted lines are guides to the eye only.
    b) Normalized surface tension $\gamma/\gamma_0$ as a function of reduced temperature, $\tau_\text{I}$. 
    c) Critical amplitude $\gamma_0$ as a function of the polar fraction, $f_\text{P}$, of the different systems. The dotted line is a guide to the eye only.
    d) Difference of $\gamma$ between the homopolymer and heteropolymer systems, $\Delta \gamma = \gamma_{\text{H}_{60}} - \gamma_{\text{H}_n\text{P}}$, as a function of $\tau_\text{I}$. Open symbols show simulation results, while dotted lines show the best fit to $\Delta \gamma = \Delta \gamma_0 \tau_\text{I}^2$.} 
    \label{fig:surface.tension}
\end{figure}

We measured the surface tension $\gamma$ in our slab simulations from the anisotropy of the pressure tensor through the Kirkwood-Buff expression (see Methods Section). All four systems exhibit the same expected trend, where $\gamma$ decreases as a function of temperature, approaching $\gamma \to 0$ at the system's respective critical temperature, where the interface itself disappears (Fig.~\ref{fig:surface.tension}a). 

Close to the critical point, thermodynamic variables exhibit a power-law dependence with respect to $\tau$, e.g. the surface tension scales as
\begin{equation}
    \gamma(\tau) = \gamma_0 \tau_\text{I}^\mu,
    \label{eq.gamma.scaling}
\end{equation}
where $\gamma_0$ is the system-specific critical amplitude, $\tau_\text{I}$ is the reduced temperature based on $T_\text{c,I}$, and $\mu \approx 1.26$ is the critical exponent of the three-dimensional Ising universality class \cite{widom:jcp:1965}. Because $\mu$ is universal, $\gamma_0$ isolates the sequence- and composition-dependent contribution to the interfacial free energy near criticality.

Figure~\ref{fig:surface.tension}b shows the normalized surface tension,  $\gamma/\gamma_0$, as a function of $\tau_\text{I}$. Once rescaled by $\gamma_0$, the data from all systems collapse onto the single Ising scaling relation $\gamma/\gamma_0 = \tau_\text{I}^\mu$, confirming that they share the same universal critical behavior. The collapse holds up to $\tau_\text{I} \simeq 0.5$; the deviations at larger $\tau_\text{I}$ mark the edge of the asymptotic critical regime, where the systems deviate from the power-law behavior. The composition-dependent  differences in surface tension are thus captured by the critical amplitude, $\gamma_0$, which decreases systematically with increasing polar fraction, $f_\text{P}$ (Figure~\ref{fig:surface.tension}c). Regular
repeating sequences with higher polar content therefore exhibit systematically lower surface tension at matched $\tau$. For the $[\text{H}_{2}\text{P}]_{20}$ system, the amplitude is reduced to $\gamma_{0, \text{H}_{2}\text{P}} / \gamma_{0, \text{H}_{60}} \approx 0.55$, that is about half that of the purely hydrophobic system. Thus, after accounting for the distance from criticality, the surface tension can still differ by nearly a factor of two, reflecting the composition-dependence of the interfacial free energy.

To quantify this composition-induced reduction directly, we compare each heteropolymer system to the purely hydrophobic reference system, $[\text{H}_{60}]_1$, defining
\begin{equation}
    \Delta \gamma(\tau_\text{I}) = \gamma_{\text{H}_{60}}(\tau_\text{I}) - \gamma_{\text{H}_n\text{P}}(\tau_\text{I}).
\end{equation}
Away from the asymptotical critical regime, $\tau_\text{I} \gtrsim 0.3$, the $\Delta \gamma$ data for all three heteropolymer systems can be described by a quadratic form,
\begin{equation}
    \Delta\gamma
    =
    \Delta\gamma_0\, \tau_\text{I}^{2},
\end{equation}
with a prefactor $\Delta\gamma_0$ that increases with polar fraction. The surface-tension reduction therefore becomes more pronounced both as polar content increases and as the system moves further from its critical point.

A similar quadratic dependence was predicted theoretically for a single heteropolymer adsorbed at a selective interface \cite{sommer:epl:1995}, where the free energy gain of HP subunits at the interface grows quadratically with the selective strength between the two coexisting phases. This emerges from the physical analogy between a diblock sequence located at the interface between two coexisting phases with opposite selectivities for the two residue types and an induced dipole in an external field \cite{sommer:mm:1994}. The observed dependence is consistent with self-surfactancy: polar residues, or polar-rich chain segments, preferentially orient toward the dense--dilute interface, reducing the energetically unfavorable exposure of hydrophobic residues to the surrounding aqueous phase and thereby lowering the interfacial tension.


\subsection{Structure}
\label{Sec.Results.Structure}

\begin{figure*}[t]
    \centering
    \includegraphics[width=\textwidth]{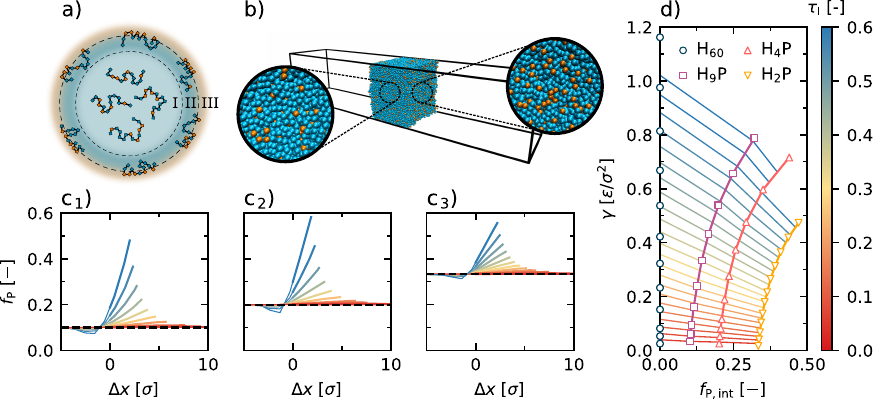}
    \caption{Accumulation of polar residues at the interface.
    a) Schematic illustration of the structural arrangement of heteropolymers at the interface, with residues distributed homogeneously in the dense phase far from the interface (I), slight excess of hydrophobic residues on the dense side of the interface (II), and a marked excess of polar residues on the dilute side of the interface (III). 
    b) Simulation snapshot of the $[\text{H}_9\text{P}]_6$ system at $\tau_\text{I} \simeq 0.64$, with magnified views highlighting the distribution of residues within the dense phase (left) and at the interface (right).
   c) Fraction of polar residues $f_\text{P}(\Delta x)$ as a function of the distance to the interface center, $\Delta x$, averaged over the two interfaces, for the $[\text{H}_9\text{P}]_6$ ($\text{c}_1$), $[\text{H}_4\text{P}]_{12}$ ($\text{c}_2$) and $[\text{H}_2\text{P}]_{20}$ systems ($\text{c}_3$). Dashed black lines correspond to the $f_\text{P}$ bulk composition. Results shown for various reduced temperatures $\tau_\text{I}$, as indicated in the color bar. The thickness of each line is increased in the interval $\Delta x =$[0, $2d$].
    d) The surface tension, $\gamma$, as a function of the accumulation of polar residues in the interval $\Delta x =$[0, $2d$], $f_\text{P, int}$. Symbols correspond to the simulation results, while colored lines indicate linearly interpolated values at matched $\tau_\text{I}$. 
    } 
    \label{fig:structure.regular}
\end{figure*}

We hypothesize that the decrease in surface tension originates from HP-subunits that orient with respect to the interface, leading to an enhanced polar fraction on the dilute side of the interface, while the hydrophobic residues remain preferentially localized on the dense side of the interface (Fig.~\ref{fig:structure.regular}a). We test this hypothesis by determining the local residue composition across the interface and then ask whether the measured degree of polar enrichment correlates with the surface tension.

The accumulation of polar residues at the dense--dilute interface can be seen already from the simulation snapshots (Fig.~\ref{fig:structure.regular}b). We quantify the enrichment of polar residues at the interface by computing the local fraction of polar residues, $f_\text{P}(\Delta x)$, as a function of the distance from the interface center, $\Delta x$ (Fig.~\ref{fig:structure.regular}c). Deep in the dense phase ($\Delta x < 0$), $f_\text{P}(\Delta x)$ matches the system's bulk composition $f_\text{P}$. Approaching the interface from the dense side, $f_\text{P}(\Delta x)$ first dips slightly below the bulk value, before rising above it on the dilute side of the interface. The  $f_\text{P}(\Delta x)$ profiles therefore reveal the predicted interfacial rearrangement: hydrophobic residues are preferentially retained on the dense side, whereas polar residues become enriched toward the dilute side. This inhomogeneity is observed for all three regular sequences, although its magnitude depends on the system.

The enrichment of polar residues at the interface provides a natural explanation for the reduction in surface tension: because some residues must inevitably occupy the interface, it is energetically favorable for polar residues to do so, as it maximizes the number of attractive contacts between hydrophobic residues in the dense phase. If this mechanism is responsible for the reduced surface tension, then $\gamma$ should correlate directly with the local polar fraction at the outer side of the interface, rather than with the bulk composition alone. To test this hypothesis, we compute the average polar fraction on the dilute side of the interface, $f_\text{P, int}$, as

\begin{equation}
    f_\text{P, int} = \frac{1}{2d} \int_{0}^{2d} f_\text{P}(\Delta x)\, d\Delta x,
    \label{eq.fP.Int}
\end{equation}

where the center of the interface $x_0$ and the interfacial width $d$ are both extracted from the hyperbolic-tangent fit to the density profile (see Methods Section).

Figure~\ref{fig:structure.regular}$\text{d}$ shows $\gamma$ as a function of $f_\text{P,int}$ for  sequences of type $\text{H}_n\text{P}$. Because the simulations were performed at matched absolute temperature rather than matched $\tau$, we linearly interpolate between neighboring data points to compare systems at matched $\tau_\text{I}$ (colored lines). Across all sequences, $\gamma$ decreases monotonically with increasing $f_\text{P,int}$, providing direct evidence that interfacial polar enrichment is associated with lower surface tension.


\subsection{Sequence modulations}
\label{Sec.Results.SequenceModulations}

\begin{figure*}[t]
    \centering
    \includegraphics[width=\textwidth]{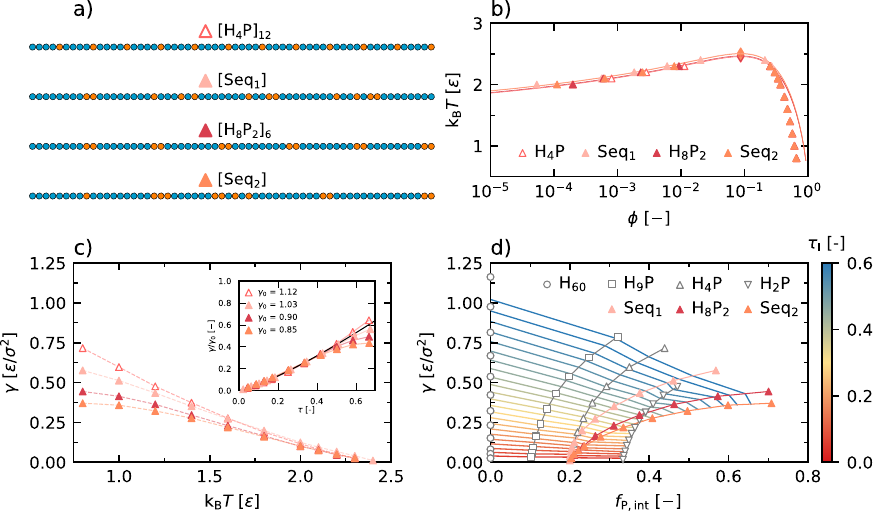}
    \caption{Effects of sequence modulations on the phase behavior and surface tension.
    a) Four heteropolymers with equal composition $f_\text{P}=1/5$ but different sequences. Hydrophobic and polar residues are shown in blue and orange, respectively.
    b) Binodals of the four systems in absolute temperature. The statistical uncertainties of the data are smaller than the symbol size. Solid lines are fits to the standard FH model, while diamonds mark the critical temperatures obtained from the FH fit.
    c) Surface tension, $\gamma$, as a function of the absolute temperature, $\text{k}_\text{B}T$. Open symbols show $\gamma$ from simulations, while and dashed lines are guides to the eye only. The insert shows the rescaled surface tension, $\gamma/\gamma_0$, as a function of the reduced temperature, $\tau_\text{I}$, with the respective critical amplitudes $\gamma_0$ indicated in the legend. 
    d) $\gamma$ as a function of the accumulation of polar residues in the interval $[x_0, x_0+2d]$, $f_\text{P, int}$ for all systems. Symbols correspond to the simulation measurements, while colored lines connecting different systems indicate linearly interpolated values at matched $\tau_\text{I}$. } 
    \label{fig:sequence.modulations}
\end{figure*}

So far, we have focused on regular, repeating sequences of the type $\text{H}_n\text{P}$, for which we established two central results: (i) their phase behavior is well described by the homopolymer FH model, with the effective interaction parameter $\Delta\varepsilon$ decreasing with increasing polar fraction $f_\text{P}$, and (ii) their surface tension decreases systematically with increasing polar fraction at matched thermodynamic states. To probe the effects of sequence irregularity, as commonly observed in natural IDPs, we next test to what extent these relationships persist under sequence variation at fixed composition. We repeat our analysis for a set of systems that share the same composition, $f_\text{P}=1/5$, but differ in their sequence: the regular sequence $[\text{H}_4\text{P}]_{12}$ studied above, a blockier sequence $[\text{H}_8\text{P}_2]_6$, and two irregular sequences, $\text{Seq}_1$ and $\text{Seq}_2$, see Fig. \ref{fig:sequence.modulations}a. For each system, we simulate a single fixed sequence, mimicking a condensate composed of a single protein species.

The binodals of all four sequences are well described by the FH model. Surprisingly, the individual binodals are nearly indistinguishable, exhibiting critical temperatures within  $0.03\,\text{k}_\text{B}T$ (about 1\%) of one another (Fig.~\ref{fig:sequence.modulations}b). Thus, over the sequence modulations considered here, phase behavior is governed almost entirely by the overall composition of polar residues $f_\text{P}$. This insensitivity is, however, not general. Blocky sequences with a high polar fraction can lose the homogeneous dense phase, forming first structured liquids with microphase-separated polar regions and then finite-size micelles \cite{rekhi:jpcb:2023}. This transition occurs at considerably smaller $f_\text{P}$ when H--P contacts are purely repulsive \cite{statt:jcp:2020, rekhi:jpcb:2023}.

Although the coexistence densities are almost indistinguishable for our sequences, the surface tension shows large deviations. Because all four $f_\text{P}=1/5$ systems share nearly the same critical temperature, their surface-tension curves approach zero at approximately the same absolute temperature. At lower temperatures, however, the system's surface tension differs substantially (Fig.~\ref{fig:sequence.modulations}c). Importantly, the sequence modulations do not affect the critical behavior of the surface tension close to criticality, where each system follows the predicted scaling of the three-dimensional Ising universality class, $\gamma \sim \gamma_0\,\tau^\mu$ (inset of Fig.~\ref{fig:sequence.modulations}c), with distinct critical amplitudes $\gamma_0$, revealing the extent of self-surfactancy of the respective sequence. Quantitatively, the critical amplitudes reveal a surface tension reduction of the $\text{Seq}_2$ system of $\sim25\%$ compared to the $[\text{H}_4\text{P}]_{12}$ system at matched thermodynamic conditions within the critical regime. At even lower temperatures, away from the asymptotic critical regime, the difference in surface tension becomes even more pronounced.

These results demonstrate that sequence, independent of composition, modulates the strength of interfacial ordering and, consequently, the surface tension. Strikingly, the surface tension of the $[\text{H}_8\text{P}_2]_6$ and $\text{Seq}_2$ systems with $f_\text{P}=1/5$ can fall below that of a system with substantially higher polar fraction ($f_\text{P}=1/3$) at matched $\tau$. Thus, while sequence has little discernible effect on phase behavior, it can substantially alter the surface tension of the resulting condensates. To understand the structure-property relation of these sequences, we compute the average polar fraction on the dilute side of the interface, $f_\text{P,int}$, for each sequence (Fig.~\ref{fig:sequence.modulations}d). Notably, the temperature dependence of interfacial polar enrichment differs between sequences, yet when compared at matched reduced temperature, $\tau_\text{I}$, the amount of polar residues accumulated at the interface consistently correlates with a reduction in surface tension across all systems. Number density profiles of terminal residues show that chain ends generally accumulate at the interface, with polar ends exhibiting particularly strong enrichment (Fig. S8--S9 in the SI). This accumulation is consistent with the conformational entropy gained by localizing chain ends at the interface \cite{wang:ml:2023, bauer:nc:2024}.

It is known from studies of copolymer adsorption at selective interfaces that random copolymers display stronger and more robust interfacial localization as compared to short repeating blocks, driven by the random excess in the local composition along a given sequence \cite{sommer:epl:1995}. To better understand such sequence effects, we constructed 12 additional disordered sequences at fixed $f_\text{P} = 1/5$. Here, we determined $\gamma$ at a single absolute temperature $k_\text{B}T=0.8\,\varepsilon$, since constructing the entire coexistence curve for each additional sequence is computationally demanding. The miscibility gap, $\rho_\text{l} - \rho_\text{v}$, of these additional sequences lies within the previously established ones of $[\text{H}_4\text{P}]_{20}$ and $\text{Seq}_2$ (Fig. S13a in the SI), indicating similar critical temperatures $T_\text{c}$.  Across this larger sequence library, $\gamma$ tends to decrease with increasing local clustering of P residues, although considerable sequence-to-sequence variation remains (see Sec.~S7 in the SI). Barring the potential change in $\tau_\text{I}$, we find that $\gamma$ varies by a factor of $2.6$ for the investigated sequences at fixed composition $f_\text{P} = 1/5$ at $\text{k}_\text{B}T = 0.8 \varepsilon$, far below the critical temperature.


\section{Discussion}
\label{Sec.Discussion}

Biomolecular condensates form interfaces whose mechanical properties -- such as those governing fusion, wetting, and coarsening -- emerge from a combination of molecular features and thermodynamic conditions, such as protein sequence and valency, composition stoichiometry, temperature, and pH, yet in experiments these contributions are hard to disentangle. Simulations, by contrast, allow the influence of specific microscopic features, such as the amphiphilic character of the chain, to be isolated and studied in a controlled way. Here, we ask whether the amphiphilicity encoded by protein sequence can control interfacial tension by itself, through the preferential partitioning of polar residues to the interface.

At matched distance from the critical point, we find that composition and sequence jointly determine the value of the surface tension through ``self-surfactancy'' -- a mechanism in which the phase-separated material supplies its own amphiphile. Energetically, it is beneficial to promote polar residues to the interface because it reduces the unfavorable exposure of hydrophobic residues to the solvent. Systems with higher polar content can promote more residues to the interface, resulting in lower surface tensions. This energetic drive competes with a configurational cost that depends on sequence, since a polar residue flanked by hydrophobic neighbors cannot move to the surface without dragging unfavorable contacts along. Thus, patterns that lower this cost can make self-surfactancy more efficient, and further reduce surface tension at the same composition. Crucially, for the present model and investigated sequences, bulk phase behavior is controlled almost entirely by overall composition, and is well described by the Flory-Huggins theory. Sequence therefore acts as an independent control knob for surface tension while leaving bulk phase behavior essentially unchanged.

In cells, where condensates are often multiphasic and their morphology is set by the relative interfacial tensions between coexisting phases, self-surfactancy could allow sequence to tune their wetting behavior and coarsening. Furthermore, cells are known to regulate interfacial tension through dedicated surfactant-like proteins; self-surfactancy would be a complementary mechanism that needs no additional component, and one that mutations could in principle tune. We also see a design principle for synthetic condensates: given a fixed composition, the patterning of polar and hydrophobic residues could set interfacial properties independently of phase separation, building on sequence-based engineering of condensate wetting behavior. This is directly testable. Variants with identical composition but different patterning could be expressed or synthesized, matched in saturation concentration, and compared by micropipette aspiration or droplet fusion assays. We hope our results encourage such experiments.

We note several limitations when interpreting the results of our minimal model in the context of natural biomolecular condensates. Our model uses a two-letter alphabet, which we suspect amplifies self-surfactancy, since every residue is either strongly hydrophobic or strongly polar. Moreover, we have isolated amphiphilic effects from other microscopic interactions, such as electrostatic interactions due to charged residues. In particular, counterion localization within the condensate and at its surface can play an important role. How self-surfactancy interacts with other sequence features such as charge patterning, salt and pH, remains an open question. Finally, the investigated sequences qualitatively indicate how composition and sequence control surface tension, but the sample size is too small to identify definitive sequence features that enhance or suppress self-surfactancy.

More broadly, our results show that surface tension is not simply a by-product of phase separation, but a property that sequence can encode by its own. Sequence thus provides a handle on how condensates interact with their surroundings and with each other, one that natural evolution may already exploit and that protein engineers could deliberately design.

\section{Methods}
\label{Sec.Methods}


\subsection{Simulation methods} 
\label{Sec.Methods.SimulationModels}
We perform coarse-grained molecular dynamics (MD) simulations of phase-separating heteropolymers using a minimal hydrophobic-polar (HP) model \cite{dill:bc:1985, dill:mm:1989, dill:bc:1990, statt:jcp:2020, rekhi:jpcb:2023}. In this model, each residue is represented by a single bead of diameter $\sigma$ and mass $m$, and residues are classified as either hydrophobic (H) or polar (P). The solvent is treated implicitly, and solvent-mediated effects are incorporated through effective interactions between residues.

Non-bonded pair interactions are described by a modified Lennard-Jones (LJ) potential with a cutoff distance of $r_\text{c} = 3\,\sigma$,
\begin{equation}
    U_\text{nb}(r) = \begin{cases}
    U_\text{LJ} + \left( 1 - \lambda \right)\varepsilon, &  r <2^{1/6} \sigma,\\
    \lambda U_\text{LJ}, & r \ge  2^{1/6} \sigma,
    \end{cases}
\end{equation}
with a parameter $\lambda\in[0, 1]$ that controls the strength of the attraction. The potential interpolates between the purely repulsive Weeks-Chandler-Andersen (WCA) potential ($\lambda=0$) to the standard LJ potential ($\lambda = 1$). Following previous work \cite{rekhi:jpcb:2023}, we use $\lambda=1$ for the H--H interactions, $\lambda=0$ for the P--P interactions, and $\lambda = 1/2$ for the H--P interactions.

Chain connectivity is maintained using a finite extensible non-linear elastic (FENE) potential combined with WCA excluded volume interactions between bonded particles,
\begin{equation}
    U_\text{b}(r) = U_\text{WCA} - \frac{1}{2} K r_\text{0}^2\ \ln \left[ 1 - \left( \frac{r}{r_\text{0}}\right)^2\right],
\end{equation}
with spring constant $K = 30\,\varepsilon/\sigma^2$ and maximum bond extension $r_0 = 1.5\, \sigma$. We use reduced LJ units, setting $\sigma$, $m$ and $\varepsilon$ as the units of length, mass and energy, respectively. The corresponding MD time unit is $\tau_\text{MD} = \sigma \sqrt{m/\varepsilon}$. 

Each system contains $N=500$ linear chains of length $M=60$. The slab simulations are performed in simulation boxes of dimensions $L_x\ =\ 176.0\ \sigma$ and $L_y\ =\ L_z\ =\ 30.0\ \sigma$, with periodic boundary conditions applied to all three directions.

The simulations are performed using HOOMD-blue (v.5.1.0) \cite{anderson:cms:2020}. The equations of motion are integrated using the velocity-Verlet algorithm with a time step of $\Delta t = 0.002 \tau_\text{MD}$. Temperature is controlled using a Langevin thermostat with friction coefficient $m/\tau_\text{MD}$.

For each sequence, initial single-chain simulations were first performed to generate sample chain conformations. Configurations extracted from these single-chain simulations were then used to initialize $N=500$ chains on a simple cubic lattice at an initial monomer number density of $\rho \approx 0.013\,\sigma^{-3}$. The system was subsequently compressed to a simulation box of size $27.0\,\times 30.0 \times 30.0\,\sigma^3$. During this compression, the number density was increased linearly in time while maintaining the cubic box shape. The resulting dense bulk conformations were then embedded into an elongated simulation box of size $L_x =176.0\, \sigma$ and $L_y = L_z = 30.0\,\sigma$, thereby generating liquid slabs surrounded by dilute vapor. Particles crossing the periodic boundary in the original dense bulk configuration were re-wrapped to preserve chain connectivity after insertion into the elongated box.

Each slab simulation was equilibrated until the monomer density profile along the $x$-direction reached a stationary form. Depending on temperature and sequence, equilibration required between $2 \times 10^5\,\tau_\text{MD}$ and $5 \times 10^5\, \tau_\text{MD}$. Production runs were then performed for $2 \times 10^6\, \tau_\text{MD}$. During the production runs, full configurations were saved every  $2 \times 10^3\,\tau_\text{MD}$ and the global pressure tensor was sampled every $2\,\tau_\text{MD}$. The parameters of all investigated systems are listed in Tables S3-S9 of the SI.


\subsection{Phase behavior} 
\label{Sec.Methods.PhaseBehavior}
For each slab simulation, we computed the time-averaged monomer number density profile, $\rho(x)$, along the slab-normal direction, $x$. Prior to averaging, each saved configuration was shifted such that the center-of-mass of the dense slab coincided with the origin of the simulation box. This procedure removes translational diffusion of the dense phase along the elongated direction. The stationary monomer number density was fitted to a hyperbolic tangent function,
\begin{equation}
    \rho(x) = \frac{\rho_\text{l} + \rho_\text{v}}{2} + \frac{\rho_\text{l} - \rho_\text{v}}{2} \tanh \left( \frac{x - x_\text{0}}{d}\right),
    \label{eq:tanh}
\end{equation}
where $\rho_\text{l}$ and $\rho_\text{v}$ are the densities of the dense (``liquid'') and dilute (``vapor'') phases, respectively, $x_0$ is the center of the interface, and $d$ the interfacial width. The coexisting densities $\rho_\text{l}$ and $\rho_\text{v}$ were extracted from the plateau regions of the density profile, far from the liquid-vapor interfaces, and then converted to a volume fraction using a protein density of $1.35\,\text{g}/\text{mol}$. The values for $\rho_\text{l}$, $\rho_\text{v}$, and $d$ are listed in Tables S3-S9 of the SI.

To determine binodals within the FH model, we fit the coexisting volume fractions to the FH mean-field free energy, 
\begin{equation}
    \frac{F(\phi)}{k_\text{B}T} = \frac{1}{M_\text{eff}} \phi \log(\phi) + (1-\phi) \log(1-\phi) + \frac{\Delta \varepsilon}{k_\text{B}T}  \phi (1-\phi),
\label{eq.FH}
\end{equation}
where $\phi$ is the polymer volume fraction, $M_\text{eff}$ is the effective chain length, and $\Delta \varepsilon$ parametrizes the effective site-to-site contact energy, related to the the FH parameter through $\chi = \Delta \varepsilon / \text{k}_\text{B}T$. We optimized ($M_\text{eff}$, $\Delta \varepsilon$) for the homopolymer system, for which we have obtained $M_\text{eff} = 112.74$ and $\Delta \varepsilon = 1.895 \pm 0.003$. For the subsequent fits, we fixed the effective chain length to $M_\text{eff} = 110$ across all systems, and fitted the remaining binodals to obtain the single fit-parameter $\Delta \varepsilon$ of each of the heteropolymer systems.


\subsection{Surface tension} 
\label{Sec.Methods.SurfaceTension}

The liquid--vapor surface tension was computed from the anisotropy of the pressure tensor using the Kirkwood-Buff mechanical route. For a planar slab with surface normal along $x$, the surface tension is given by
\begin{equation}
    \gamma = \frac{1}{2} L_x \left[ \langle P_{xx}\rangle - \frac{1}{2} \left( \langle P_{yy} \rangle + \langle P_{zz} \rangle\right) \right].
\end{equation}
The factor $\frac{1}{2}$ accounts for the two liquid--vapor interfaces present in the slab geometry. Here, $P_{xx}$ is the normal component of the pressure tensor and $P_{yy}$ and $P_{zz}$ are the two tangential components. All $\gamma$ values are listed in Tables S3-S9 of the SI.

The global pressure tensor was computed by HOOMD-blue from the kinetic and virial contributions of all particles in the simulation box,
\begin{equation}
    P_{\alpha \beta}\ =\  \frac{1}{V}\ \left[ \sum_i m_i v_{i, \alpha}^2 + \sum_i W_{i, \alpha \alpha} \right],
\end{equation}
where $V=L_x L_y L_z$ is the box volume, $v_{i, \alpha}^2$ is the $\alpha$ component of the velocity of particle $i$, and $W_{i, \alpha \alpha}$ is the virial contribution associated with particle $i$. 

Surface tension uncertainties were estimated using block averaging. The time series of instantaneous surface tension estimates was divided into five blocks of equal length. The reported surface tension is the mean over the block averages, $\bar{\gamma} = \frac{1}{N_b} \sum_{k=1}^{N_b} \gamma_k$, with $N_b\ =\ 5$. Statistical uncertainties were estimated from the standard error of the block averages.

\section*{Acknowledgments}
This work was supported by the Deutsche Forschungsgemeinschaft (DFG, German Research Foundation) through the framework of the research training group RTG 3120 (Project No. 542285965), and through the Heisenberg grant to A.N. (Project No. 470113688). We further acknowledge support by the Klaus Tschira Foundation through Project No. 00.050.2024/ID 25347. The authors acknowledge the Center for High-Performance Computing (ZIH) Dresden for providing computational resources.

\section*{Conflicts of interest}
The authors have no conflicts to disclose.

\section*{Author contributions}
N. G.: Conceptualization, Data curation, Formal analysis, Investigation, Methodology, Software, Validation, Visualization, Writing - original draft;\\J.-U. S.: Conceptualization, Funding acquisition, Project administration, Resources, Writing - review \& editing;\\A. N.: Conceptualization, Funding acquisition, Project administration, Resources, Writing - original draft, Writing - review \& editing;

\section*{Data availability}
The data supporting the findings of this study will be deposited in a public repository and made available upon publication. The underlying data and analysis files can be made available to editors and reviewers during the peer-review process upon request.

\bibliography{references}

@Article{alston:jpcb:2023,
  Title                    = {The Analytical Flory Random Coil Is a Simple-to-Use Reference Model for Unfolded and Disordered Proteins},
  Author                   = {Alston, J. and Ginell, G. M. and Soranno, A. and Holehouse, A. S.},
  Journal                  = {J. Phys. Chem. B },
  Year                     = {2023},
  Pages                    = {4746--4760},
  Volume                   = {127}
}

@article{dill:bc:1985,
  title={Theory for the folding and stability of globular proteins},
  author={Dill, K. A.},
  journal={Biochemistry},
  volume={24},
  pages={1501--1509},
  year={1985}
}

@article{dill:mm:1989,
  title={A lattice statistical mechanics model of the conformational and sequence spaces of proteins},
  author={Lau, K. F. and Dill, K. A.},
  journal={Macromolecules},
  volume={22},
  pages={3986--3997},
  year={1989}
}

@article{dill:bc:1990,
  title={Dominant forces in protein folding},
  author={Dill, K. A.},
  journal={Biochemistry},
  volume={29},
  pages={7133--7155},
  year={1990}
}

@article{witzky:jcp:2025,
  title={From heteropolymer stiffness distributions to effective homopolymers. Part 1: Theoretical modeling and computational verification},
  author={Witzky, Y. and Schmid, F. and Nikoubashman, A.},
  journal={J. Chem. Phys.},
  volume={163},
  pages={164907},
  year={2025}
}

@article{witzky:jcp:2025B,
  title={From heteropolymer stiffness distributions to effective homopolymers. Part 2: Conformational analysis of intrinsically disordered proteins},
  author={Witzky, Y. and Schmid, F. and Nikoubashman, A.},
  journal={J. Chem. Phys.},
  volume={163},
  pages={164908},
  year={2025}
}

@article{brady:pnas:2017,
  title={Structural and hydrodynamic properties of an intrinsically disordered region of a germ cell-specific protein on phase separation},
  author={Brady, J. P. and Farber, P. J. and Sekhar, A. and Lin, Y.-H. and Huang, R. and Bah, A. and Nott, T. J. and Chan, H. S. and Baldwin, A. J. and Forman-Kay, J. D. and Kay, L. E.},
  journal={Proc. Natl. Acad. Sci. U. S. A.},
  volume={114},
  number={39},
  pages={E8194},
  year={2017},
  publisher={National Academy of Sciences}
}

@book{flory:cornell:1953,
  title={Principles of polymer chemistry},
  author={Flory, Paul J},
  year={1953},
  publisher={Cornell university press}
}

@article{martin:sci:2020,
  title={Valence and patterning of aromatic residues determine the phase behavior of prion-like domains},
  author={Martin, Erik W. and Holehouse, Alex S. and Peran, Ivan and Farag, Mina and Incicco, J. Jeremias and Bremer, Anne and Grace, Christy R. and Soranno, Andrea and Pappu, Rohit V. and Mittag, Tanja},
  journal={Science},
  volume={367},
  pages={694--699},
  year={2020}
}

@article{mccall:nc:2025,
  title={A label-free method for measuring the composition of multicomponent biomolecular condensates},
  author={McCall, P. M. and Kim, K. and Shevchenko, A. and Ruer-Gru{\ss}, M. and Peychl, J. and Guck, J. and Shevchenko, A. and Hyman, A. A. and Brugu{\'e}s, J.},
  journal={Nat. Chem.},
  volume={17},
  pages={1891--1902},
  year={2025}
}

@article{mitra:rpp:2026,
  title={Distinguishing near- versus off-critical phase behaviors of intrinsically disordered proteins},
  author={Mitra, G. and Ghosh, S. and Ruff, K. M. and Zhang, R. and Chauhan, G. and Pappu, R. V.},
  journal={Rep. Prog. Phys.},
  volume={89},
  pages={068101},
  year={2026}
}

@article{murthy:nsmb:2019,
  title={Molecular interactions underlying liquid-liquid phase separation of the FUS low-complexity domain},
  author={Murthy, A. C. and Dignon, G. L. and Kan, Y. and Zerze, G. H. and Parekh, S. H. and Mittal, J. and Fawzi, N. L.},
  journal={Nat. Struct. Mol. Biol.},
  volume={26},
  pages={637--648},
  year={2019}
}

@article{schuster:nc:2018,
  title={Controllable protein phase separation and modular recruitment to form responsive membraneless organelles},
  author={Schuster, B. S. and Reed, E. H. and Parthasarathy, P. and Jahnke, C. N. and Caldwell, R. M. and Bermudez, J. G. and Ramage, H. and Good, M. C. and Hammer, D. A.},
  journal={Nat. Commun.},
  volume={9},
  pages={2985},
  year={2018}
}

@article{qian:acs:2022,
  title={Analytical solution to the Flory--Huggins model},
  author={Qian, Daoyuan and Michaels, Thomas C. T. and Knowles, Tuomas P. J.},
  journal={J. Phys. Chem. Lett.},
  volume={13},
  number={33},
  pages={7853--7860},
  year={2022},
  publisher={ACS Publications}
}

@Article{bauer:nc:2024,
  Author = {Bauer, D. J. and Nikoubashman, A.},
  Journal = {Nat. Commun.},
  Pages = {9975},
  Title = {The conformations of protein chains at the interface of biomolecular condensates},
  Volume = {15},
  Year = {2024}
}

@Article{boeddeker:np:2022,
  Author = {B{\"o}ddeker, T. J. and Rosowski, K. A. and Berchtold, D. and Emmanouilidis, L. and Han, Y. and Allain, F. H. T. and Style, R. W. and Pelkmans, L. and Dufresne, E. R.},
  Journal = {Nat. Phys.},
  Pages = {571--578},
  Title = {Non-specific adhesive forces between filaments and membraneless organelles},
  Volume = {18},
  Year = {2022}
}

@Article{farag:nc:2022,
  Author = {Farag, M. and Cohen, S. R. and Borcherds, W. M. Bremer, A. and Mittag, T. and Pappu, R. V.},
  Journal = {Nat. Commun.},
  Pages = {7722},
  Title = {Condensates formed by prion-like low-complexity domains have small-world network structures and interfaces defined by expanded conformations},
  Volume = {13},
  Year = {2022}
}

@Article{folkmann:sci:2021,
  Author = {Folkmann, A. W. and Putnam, A. and Lee, C. F. and Seydoux, G.},
  Journal = {Science},
  Pages = {1218--1224},
  Title = {Regulation of biomolecular condensates by interfacial protein clusters},
  Volume = {373},
  Year = {2021}
}

@Article{wang:ml:2023,
  Author = {Wang, Jiahui and Devarajan, Dinesh Sundaravadivelu and Nikoubashman, Arash and Mittal, Jeetain},
  Journal = {ACS Macro Lett.},
  Pages = {1472--1478},
  Title = {Conformational Properties of Polymers at Droplet Interfaces as Model Systems for Disordered Proteins},
  Volume = {12},
  Year = {2023}
}

@Article{devarajan:nc:2024,
  Title                    = {Sequence-dependent material properties of biomolecular condensates and their relation to dilute phase conformations},
  Author                   = {Devarajan, D. S. and Wang, J. and Szala-Mendyk, B. and Rekhi, S. and Kim, Y. C. and Nikoubashman, A. and Mittal, J.},
  Journal                  = {Nat. Commun.},
  Year                     = {2024},
  Pages                    = {1912},
  Volume                   = {15}
}

@article{elbaum-garfinkle:pnas:2015,
  title={The disordered P granule protein LAF-1 drives phase separation into droplets with tunable viscosity and dynamics},
  author={Elbaum-Garfinkle, S. and Kim, Y. and Szczepaniak, K. and Chen, C. C.-H. and Eckmann, C. R. and Myong, S. and Brangwynne, C. P.},
  journal={Proc. Natl. Acad. Sci. U. S. A.},
  volume={112},
  number={23},
  pages={7189--7194},
  year={2015},
  publisher={National Academy of Sciences}
}

@article{fisher:nc:2020,
  title={Tunable multiphase dynamics of arginine and lysine liquid condensates},
  author={Fisher, R. S. and Elbaum-Garfinkle, S.},
  journal={Nat. Commun.},
  volume={11},
  pages={4628},
  year={2020}
}

@article{ghosh:nc:2021,
  title={Shear relaxation governs fusion dynamics of biomolecular condensates},
  author={Ghosh, A. and Kota, D. and Zhou, H.-X.},
  journal={Nat. Commun.},
  volume={12},
  pages={5995},
  year={2021}
}

@article{golani:cp:2025,
  title={Mesoscale properties of protein clusters determine the size and nature of liquid-liquid phase separation (LLPS)},
  author={Golani, G. and Seal, M. and Kar, M. and Hyman, A. A. and Goldfarb, D. and Safran, S.},
  journal={Commun. Phys.},
  volume={8},
  pages={226},
  year={2025}
}

@article{holland:sm:2023,
  title={Surface tension measurement and calculation of model biomolecular condensates},
  author={Holland, J. and Castrej{\'o}n-Pita, A. A. and Tuinier, R. and Aarts, D. G. A. L. and Nott, T. J.},
  journal={Soft Matter},
  volume={19},
  number={45},
  pages={8706--8716},
  year={2023}
}

@article{legrand:jcis:1969,
  title={The Molecular Weight Dependence of Polymer Surface Tension},
  author={Legrand, D. G. and Gaines Jr., G. L.},
  journal={J. Colloid Interface Sci.},
  volume={31},
  number={2},
  pages={162--167},
  year={1969}
}

@article{pyo:pnas:2023,
  title={Proximity to criticality predicts surface properties of biomolecular condensates},
  author={Pyo, Andrew G. T. and Zhang, Yaojun and Wingreen, Ned S.},
  journal={Proc. Natl. Acad. Sci. U. S. A.},
  volume={120},
  number={23},
  pages={e2220014120},
  year={2023},
  publisher={National Academy of Sciences}
}

@article{shinozaki:poly:1982A,
  title={Interfacial tension of demixed polymer solutions near the critical temperature: polystyrene + methylcyclohexane},
  author={Shinozaki, K. and Abe, M. and Nose, T.},
  journal={Polymers},
  volume={23},
  number={5},
  pages={722--727},
  year={1982}
}

@article{shinozaki:poly:1982B,
  title={Interfacial tension of demixed polymer solutions near the critical temperature: polystyrene + methylcyclohexane},
  author={Shinozaki, K. and van Tan, T. and Saito, Y. and Nose, T.},
  journal={Polymers},
  volume={23},
  number={5},
  pages={728--734},
  year={1982}
}

@article{sommer:mm:1994,
  title={On the Adsorption Threshold of Multiblock Copolymers },
  author={Sommer, J.-U. and Halperin, A. Daoud, M.},
  journal={Macromolecules},
  volume={27},
  number={23},
  pages={6991--6992},
  year={1994}
}

@article{sommer:epl:1995,
  title={Copolymers at selective interfaces},
  author={Sommer, J.-U. and Daoud, M.},
  journal={EPL (Europhysics Letters)},
  volume={32},
  number={5},
  pages={407--412},
  year={1995}
}

@article{wang:bpr:2021,
  title={Surface tension and viscosity of protein condensates quantified by micropipette aspiration},
  author={Wang, H. and Kelley, F. M. and Milovanovic, D. and Schuster, B. S. and Shi, Z.},
  journal={Biophys. Rep.},
  volume={1},
  number={1},
  pages={100011},
  year={2021}
}

@article{widom:jcp:1965,
  title={Surface tension and molecular correlations near the critical point},
  author={Widom, Benjamin},
  journal={J. Chem. Phys.},
  volume={43},
  number={11},
  pages={3892--3897},
  year={1965},
  publisher={American Institute of Physics}
}

@article{wang:cell:2018,
  title={A molecular grammar governing the driving forces for phase separation of prion-like RNA binding proteins},
  author={Wang, Jie and Choi, Jeong-Mo and Holehouse, Alex S and Lee, Hyun O and Zhang, Xiaojie and Jahnel, Marcus and Maharana, Shovamayee and Lemaitre, R{\'e}gis and Pozniakovsky, Andrei and Drechsel, David and others},
  journal={Cell},
  volume={174},
  number={3},
  pages={688--699},
  year={2018},
  publisher={Elsevier}
}

@article{pyo:iscience:2022,
  title={Surface tension and super-stoichiometric surface enrichment in two-component biomolecular condensates},
  author={Pyo, Andrew G T and Zhang, Yaojun and Wingreen, Ned S},
  journal={Iscience},
  volume={25},
  number={2},
  year={2022},
  publisher={Elsevier}
}

@article{das:pnas:2013,
  title={Conformations of intrinsically disordered proteins are influenced by linear sequence distributions of oppositely charged residues},
  author={Das, Rahul K and Pappu, Rohit V},
  journal={Proc. Natl. Acad. Sci. U. S. A.},
  volume={110},
  number={33},
  pages={13392--13397},
  year={2013},
  publisher={National Academy of Sciences}
}

@book{israelachvili:ap:2011,
  title={Intermolecular and surface forces},
  author={Israelachvili, Jacob N},
  year={2011},
  publisher={Academic press}
}

@article{banani:nrmcb:2017,
  title={Biomolecular condensates: organizers of cellular biochemistry},
  author={Banani, S. F. and Lee, H. O. and Hyman, A. A. and Rosen, M. K.},
  journal={Nat. Rev. Mol. Cell. Biol.},
  volume={18},
  number={5},
  pages={285--298},
  year={2017}
}

@article{brangwynne:sci:2009,
  title={Germline P Granules Are Liquid Droplets That Localize by Controlled Dissolution/Condensation},
  author={Brangwynne, C. P. and Eckmann, C. R. and Courson, D. S. and Rybarska, A. and Hoege, C. and Gharakhani, J. and J{\"u}licher, F. and Hyman, A. A.},
  journal={Science},
  volume={324},
  number={5935},
  pages={1729--1732},
  year={2009}
}

@article{hyman:arc:2014,
  title={Liquid-Liquid Phase Separation in Biology},
  author={Hyman, A. A. and Weber, C. A. and J{\"u}licher, F.},
  journal={Annu. Rev. Cell Dev. Biol.},
  volume={30},
  pages={39--58},
  year={2014}
}

@article{franzmann:csp:2019,
  title={Protein phase separation as a stress survival strategy},
  author={Franzmann, Titus M and Alberti, Simon},
  journal={Cold Spring Harbor perspectives in biology},
  volume={11},
  number={6},
  pages={a034058},
  year={2019},
  publisher={Cold Spring Harbor Laboratory Press}
}

@article{su:sci:2016,
  title={Phase separation of signaling molecules promotes T cell receptor signal transduction},
  author={Su, X. and Ditlev, J. A. and Hui, E. and Xing, W. and Banjade, S. and Okrut, J. and King, D. S. and Taunton, J. and Rosen, M. K. and Vale, R. D.},
  journal={Science},
  volume={352},
  number={6285},
  pages={595--599},
  year={2016}
}

@article{sabari:sci:2018,
  title={Coactivator condensation at super-enhancers links phase separation and gene control},
  author={Sabari, Benjamin R and Dall'Agnese, Alessandra and Boija, Ann and Klein, Isaac A and Coffey, Eliot L and Shrinivas, Krishna and Abraham, Brian J and Hannett, Nancy M and Zamudio, Alicia V and Manteiga, John C and others},
  journal={Science},
  volume={361},
  number={6400},
  pages={eaar3958},
  year={2018},
  publisher={American Association for the Advancement of Science}
}

@article{hnisz:cell:2017,
  title={A phase separation model for transcriptional control},
  author={Hnisz, Denes and Shrinivas, Krishna and Young, Richard A and Chakraborty, Arup K and Sharp, Phillip A},
  journal={Cell},
  volume={169},
  number={1},
  pages={13--23},
  year={2017},
  publisher={Elsevier}
}

@article{wang:bpr:2025,
  title={The rheology and interfacial properties of biomolecular condensates},
  author={Wang, H. and Shi, Z.},
  journal={Biophys. Rev.},
  volume={17},
  pages={867--891},
  year={2025}
}

@article{anderson:cms:2020,
 Title = {HOOMD-blue: A Python package for high-performance molecular dynamics and hard particle Monte Carlo simulations},
 Author = {Joshua A. Anderson and Jens Glaser and Sharon C. Glotzer},
 Journal = {Comput. Mater. Sci.},
 Year = {2020},
 Pages = {109363},
 Volume = {173}
}

@article{rekhi:jpcb:2023,
 Title = {Role of Strong Localized vs Weak Distributed Interactions in Disordered Protein Phase Separation},
 Author = {Rekhi, S. and Devarajan, D. S. and Howard, M. P. and Kim, Y. C. and Nikoubashman, A. and Mittal, J.},
 Journal = {J. Phys. Chem. B},
 Year = {2023},
 Pages = {3829--3838},
 Volume = {127},
 Number = {17}
}

@article{schultz:jacs:1952,
  title={Phase Equilibria in Polymer-Solvent Systems},
  author={Schultz, A. R. and Flory, P. J.},
  journal={J. Am. Chem. Soc.},
  volume={74},
  number={19},
  pages={4760--4767},
  year={1952}
}

@article{silmore:mp:2017,
  title={Vapour-liquid phase equilibrium and surface tension of fully flexible Lennard–Jones chains},
  author={Silmore, Kevin S. and Howard, Michael P. and Panagiotopoulos, Athanassios Z.},
  journal={Mol. Phys.},
  volume={115},
  pages={320--327},
  year={2017}
}

@article{statt:jcp:2020,
  title={Model for disordered proteins with strongly sequence-dependent liquid phase behavior},
  author={Statt, Antonia and Casademunt, Helena and Brangwynne, Clifford P and Panagiotopoulos, Athanassios Z},
  journal={J. Chem. Phys.},
  volume={152},
  number={7},
  year={2020}
}

@article{urry:1992,
  title={Hydrophobicity scale for proteins based on inverse temperature transitions},
  author={Urry, Dan W and Gowda, D Channe and Parker, Timothy M and Luan, Chi-Hao and Reid, Michael C and Harris, Cynthia M and Pattanaik, Asima and Harris, R Dean},
  journal={Biopolymers: Original Research on Biomolecules},
  volume={32},
  number={9},
  pages={1243--1250},
  year={1992},
  publisher={Wiley Online Library}
}

\end{document}


\title{Supporting Information: Sequence-encoded amphiphilicity controls the interfacial tension of biomolecular condensates}

\author{Nora-Elen Giesinger}
\affiliation{Leibniz-Institut f{\"u}r Polymerforschung Dresden e.V., Hohe Stra{\ss}e 6, 01069 Dresden, Germany}
\affiliation{Institut f{\"u}r Theoretische Physik, Technische Universit{\"a}t Dresden, 01069 Dresden, Germany}

\author{Jens-Uwe Sommer}
\email{sommer@ipfdd.de}
\affiliation{Leibniz-Institut f{\"u}r Polymerforschung Dresden e.V., Hohe Stra{\ss}e 6, 01069 Dresden, Germany}
\affiliation{Institut f{\"u}r Theoretische Physik, Technische Universit{\"a}t Dresden, 01069 Dresden, Germany}
\affiliation{Cluster of Excellence Physics of Life, Technische Universit{\"a}t Dresden, 01062 Dresden, Germany}

\author{Arash Nikoubashman}
\email{anikouba@ipfdd.de}
\affiliation{Leibniz-Institut f{\"u}r Polymerforschung Dresden e.V., Hohe Stra{\ss}e 6, 01069 Dresden, Germany}
\affiliation{Institut f{\"u}r Theoretische Physik, Technische Universit{\"a}t Dresden, 01069 Dresden, Germany}
\affiliation{Cluster of Excellence Physics of Life, Technische Universit{\"a}t Dresden, 01062 Dresden, Germany}

\date{\today}

\maketitle

\setlength{\parindent}{0pt}
\setlength{\parskip}{0.2\baselineskip}

\section{Comparison with experiments}
Despite the importance of interfacial properties for the behavior of biomolecular condensates, surprisingly few experimental data are available on their phase coexistence behavior, particularly their critical temperatures, and surface tensions. We identified relevant data for the RGG-RGG tandem sequence of LAF-1\cite{wang:bpr:2021, schuster:nc:2018} and for two Ddx4 sequences, Ddx4(1-229) and Ddx4(1-231).\cite{holland:sm:2023} In contrast, the critical temperature and surface tensions near the critical point of classical synthetic polymer solutions are precisely characterized, and we picked polystyrene (PS) in methylcyclohexane as a reference system.\cite{shinozaki:poly:1982A, shinozaki:poly:1982B}

We first consider the RGG-RGG tandem sequence of LAF-1. Neglecting the molecular weight of the short linker between the two 168-residue RGG sequences, the molecular weight is estimated as $M_\text{w} \approx 36.9\,\text{kg}/\text{mol}$. In a buffer containing $150\,\text{mM}$ NaCl, the surface tension was report as $\gamma = 0.159 \pm 0.010\,\text{mN}/\text{m}$ from micropipette aspiration.\cite{wang:bpr:2021} Based on fusion experiments, the surface tension was estimated to lie within the range of $0.044\,\text{mN}/\text{m} < 0.257\,\text{mN}/\text{m}$.\cite{wang:bpr:2021} The temperature of these measurements was not specified precisely, and we assume that these were taken at room temperature. The critical temperature in a similar buffer containing $150\,\text{mM}$ NaCl was estimated as $T_\text{c} \approx 310\,\text{K}$ using a spectrophotometric turbidity assay, at a fixed protein concentration of $12\,\mu\text{M}$.\cite{schuster:nc:2018} The saturation concentration was determined to be approximately $1\,\mu\text{M}$ ($\approx 0.037\,\text{mg}/\text{mL}$) based on the presence of optically resolvable droplets.\cite{schuster:nc:2018} Assuming room temperature to be $T = 298\,\text{K}$, then we obtain a reduced temperature of $\tau \approx 0.04$. 

For both Ddx4(1-229) ($M_\text{w} \approx 25.3\,\text{kg}/\text{mol}$) and Ddx4(1-231) ($M_\text{w} \approx 25.6\,\text{kg}/\text{mol}$), the surface tension was determined from sessile droplets in a $150\,\text{mM}$ NaCl buffer at $T = 295\,\text{K}$. The resulting values were $\gamma \approx 0.082 \pm 0.010\,\text{mN}/\text{m}$ and $\gamma \approx 0.038 \pm 0.003\,\text{mN}/\text{m}$, respectively.\cite{holland:sm:2023} The protein concentrations in the dilute phase were measured via spectrophotometry, yielding $\rho_\text{v} \approx 27 \pm 2\,\mu\text{M}$ and $\rho_\text{v} \approx 48 \pm 2\,\mu\text{M}$ for Ddx4(1-229) and Ddx4(1-231), respectively. The concentrations of the corresponding dense phase were inferred from density calculations as $\rho_\text{l} \approx 15.9 \pm 1.8\,\text{mM}$ and $11.3 \pm 0.8\,\text{mM}$, respectively.\cite{holland:sm:2023} To estimate the distance from criticality, we use the Flory-Huggins (FH) interaction parameters $\chi$ determined by Holland et al. from the coexistence densities at $T = 295\,\text{K}$, yielding $\chi = 0.641$ for Ddx4(1-229) and $\chi = 0.616$ for Ddx4(1-231).\cite{holland:sm:2023} Comparing these values with the theoretically predicted critical interaction strength $\chi_\text{c} = 0.565$, then gives reduced temperatures of $\tau \approx 0.12$ and $\tau \approx 0.08$, respectively.

\begin{figure}[htb]
    \centering
    \includegraphics[width=10cm]{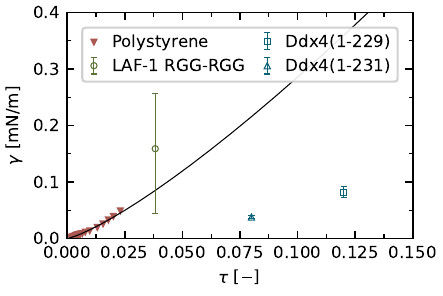}
    \caption{Experimentally measured surface tension $\gamma$ of three intrinsically disordered proteins and one synthetic polymer against the distance to the critical temperature, $\tau$. The solid line shows the Ising fit through the polystyrene data, Eq.~\eqref{eq:gamma}.}
    \label{fig:SI.exp.comparison}
\end{figure}

For comparison, we consider polystyrene with $M_\text{w} = 37.0\,\text{kg}/\text{mol}$ in methylcyclohexane. Its molecular weight is comparable to that of the LAF-1 RGG-RGG tandem. The critical temperature was determined as $T_\text{c} = 309.65\,\text{K}$ by Shinozaki et al.\cite{shinozaki:poly:1982A, shinozaki:poly:1982B} The surface tension was determined near the critical point, up to a distance of $\tau \approx 0.02$. To facilitate comparison with the biomolecular condensates at similar reduced temperature $\tau$, we fitted their data using
\begin{equation}
    \gamma(\tau) = \gamma_0\tau^\mu
    \label{eq:gamma}
\end{equation}
with a critical exponent of $\mu = 1.26$. This procedure yields a critical amplitude of $\gamma_0 = 5.2 \pm 0.1\,\text{mN}/\text{m}$. A comparison of all the experimental data is shown in Fig.~\ref{fig:SI.exp.comparison}.

\clearpage

\section{Mapping of disordered proteins to our model}
\begin{figure}[htbp]
    \centering
    \includegraphics[width=9.3cm]{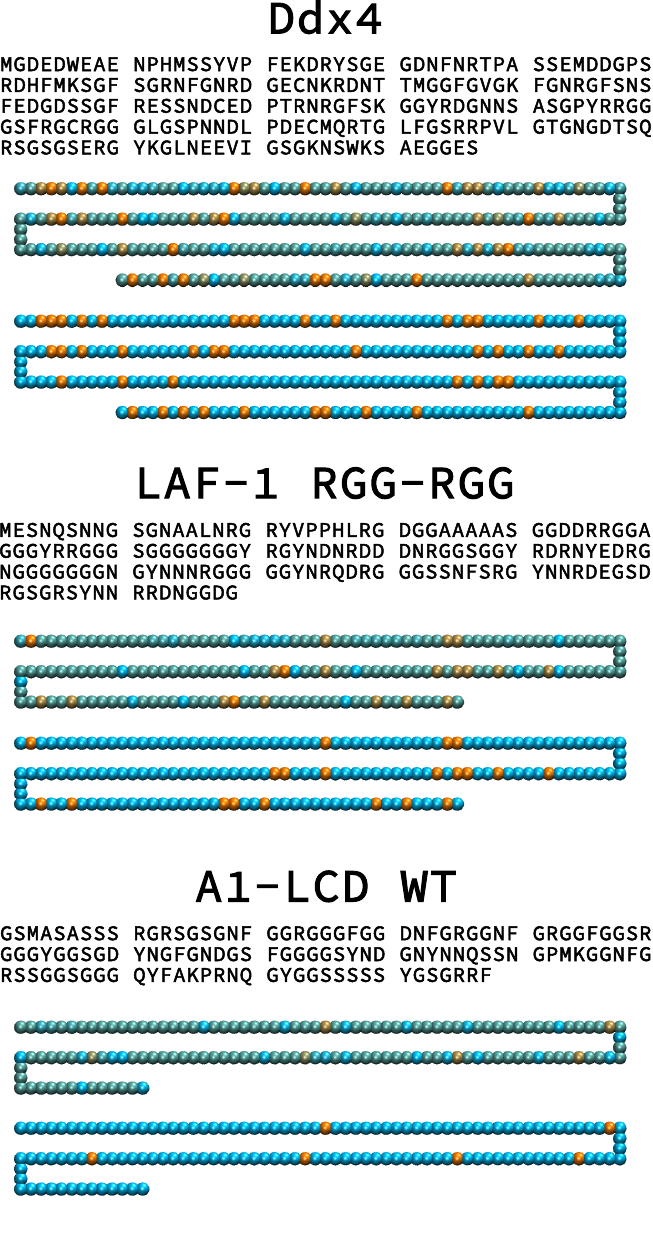}
    \caption{Illustrative mapping of the LAF-1 RGG-RGG, Ddx4, and A1-LCD WT sequences to the HP model. For each protein, the amino-acid sequence is shown alongside two one-bead-per-residue representations. In the upper representation, beads are colored by their hydrophobicity according to the Urry scale,\cite{urry:1992} with colors ranging from orange (polar, $\lambda=0$) to blue (hydrophobic, $\lambda=1.0$). We linearly interpolated the color map from orange to blue to visualize intermediate hydrophobicity values. In the lower representation, beads are classified as either polar ($\lambda \le 0.5$) or hydrophobic ($\lambda > 0.5)$. This mapping is for illustrative purposes only.}
    \label{fig:SI.CG.proteins}
\end{figure}

To illustrate the connection between amino-acid sequences with residue-specific hydrophobicities and the two-residue HP model used in this work, we visualize coarse-grained representations of three proteins: Ddx4, LAF-1 RGG-RGG and A1-LCD WT (Fig.~\ref{fig:SI.CG.proteins}). Each coarse-grained bead in the sequence corresponds to one amino acid, color coded according to its hydrophobicity, $\lambda$, using the Urry scale.\cite{urry:1992} To convert the 20-amino-acid alphabet into a two-component polymer within the HP model, we assigned residues with $\lambda \le 0.5$ to polar and residues with $\lambda > 0.5$ to hydrophobic. Using this choice of mapping, the Ddx4 protein corresponds to a polar fraction of $f_\text{P} \approx 0.19$, LAF-1 RGG-RGG to $f_\text{P} \approx 0.12$ and A1-LCD WT to $f_\text{P} \approx 0.04$. This mapping is intended only to visualize how sequence-level chemical variation can be represented in an HP model; it was not used in any simulations or subsequent analysis.

\clearpage

\section{Phase behavior and Flory-Huggins fits}
The coexisting dense and dilute phase concentrations were obtained from the plateau regions of the monomer density profiles, as described in the Methods section in the main text, and converted to volume fraction using a protein density of $\rho_\text{prt} = 1350\,\text{mg}/\text{mL}$ . For each system, the FH binodals were determined by fitting the volume fractions of the dense ($\phi_+ \equiv \rho_\text{l}/\rho_\text{prt}$) and dilute phase ($\phi_- \equiv \rho_\text{v}/\rho_\text{prt}$) to the second-order expression of the analytical solution derived in Ref.~\citenum{qian:acs:2022}:
\begin{equation}
\phi_{\pm}^{(2)}
=
\frac{
    1 - \exp\left\{
        \mp 8D \left[
            \frac{\sinh(\ln \alpha)}{\alpha^{2}}
            + D(1+\Delta)
            \left(\frac{\coth B}{\alpha^{2}}\right)
        \right]
    \right\}
}{
    1 - \exp\left\{
        \mp 8D \left[
            \left(\frac{1}{\alpha^{2}}-\alpha^{2}\right)
            \sinh \ln \alpha
            + D(1+\Delta)
            \left(
                \frac{\coth B}{\alpha^{2}}
                + \alpha^{2}\coth A
            \right)
        \right]
    \right\}
},
\label{eq.analytic.FH}
\end{equation}

where $\alpha$ and $\Delta$ are defined by the effective polymer length $M_\text{eff}$ and the FH $\chi$ parameter
\begin{equation}
    \alpha \equiv M_\text{eff}^{1/4}, \Delta \equiv \frac{\chi - \chi_\text{c}}{\chi_\text{c}},
\end{equation}
with critical interaction strength 
\begin{equation}
    \chi_\text{c} = \frac{1}{2} \left( 1 + \frac{1}{\sqrt{M_\text{eff}}} \right)^2.
\end{equation}
The parameters $A$, $B$ and $D$ are defined as
\begin{equation}
    A \equiv \frac{1}{\alpha} \left( 1 + \frac{\Delta}{\alpha^2} \right) \sqrt{3\Delta},
\end{equation}
\begin{equation}
    B \equiv \alpha \left( 1 + \alpha^2\Delta\right) \sqrt{3\Delta}
\end{equation}
and 
\begin{equation}
    D \equiv \frac{\cosh \ln \alpha}{\coth A + \coth B}. 
\end{equation}

\begin{figure}[htb]
    \centering
    \includegraphics[width=\textwidth]{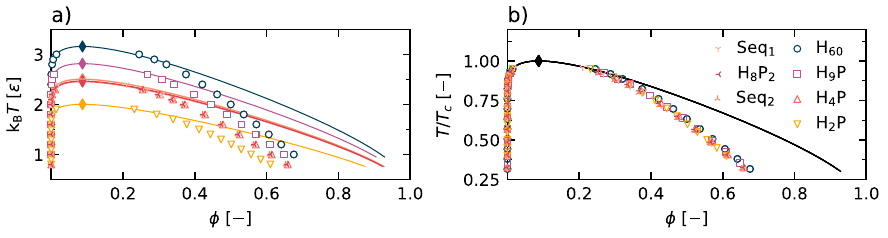}
    \caption{a) The binodals of all seven systems in absolute temperature. The statistical uncertainties of the data are smaller than the symbol size. Solid lines are fits of the standard FH model to our simulation data, and filled diamond symbols mark the critical temperatures, $T_\text{c}$, obtained from the FH fit. b) Collapse of the binodals after rescaling by $1/\text{k}_\text{B}T_\text{c}$.}
    \label{fig:SI.binodal}
\end{figure}

We optimized our simulated coexistence data to Eq.~\eqref{eq.analytic.FH} using nonlinear least squares. The sum of squared residuals was minimized in logarithmic scale, giving the dilute branch greater influence on the fit. An initial fit to the homopolymer yielded $M_\text{eff} \simeq 112.74$. For the fits reported here, $M_\text{eff}$ was fixed to $110.0$ for all sequences, including the homopolymer, leaving $\Delta \varepsilon$ as the single free fitting parameter. The resulting $\Delta \varepsilon$ values and derived FH critical temperatures, $T_\text{c}$, are summarized in Table~\ref{tab:SI.FH}, together with the coefficients of determination, $R^2$. Uncertainties in $\Delta \varepsilon$ were estimated from the least-squares Jacobian and residual variance. Figure \ref{fig:SI.binodal} shows the coexistence data and the corresponding FH fits for the homopolymer and heteropolymer systems, plotted with a linear volume-fraction axis. As is commonly observed,\cite{mitra:rpp:2026} the FH model provides a less accurate description of the dense brach, when the data are fitted with a logarithmic density scale. Nevertheless, the simulation data of the different sequences still collapse onto a single master curve after rescaling the temperature by the corresponding critical temperature, $T_\text{c}$.

\begin{table}[htb]
    \centering
    \setlength{\tabcolsep}{10pt}
    \begin{tabular}{c c c c}
        \hline
        System&  $\text{k}_\text{B} T_\text{c}$ [$\varepsilon$] &  $\Delta \varepsilon$ [$\varepsilon$] &$R^2$\\
        \hline
        $[\text{H}_{60}]_1$& 3.159 $\pm$ 0.004& 1.895 $\pm$ 0.003 &0.992\\
        $[\text{H}_{9}\text{P}]_{6}$& 2.816 $\pm$ 0.007& 1.689 $\pm$ 0.004 &0.949\\
        $[\text{H}_{4}\text{P}]_{12}$& 2.455 $\pm$ 0.004& 1.472 $\pm$ 0.003 &0.985\\
        $[\text{H}_{2}\text{P}]_{20}$& 2.002 $\pm$ 0.005& 1.201 $\pm$ 0.003 &0.991\\
        $[\text{H}_{8}\text{P}_{2}]_{6}$& 2.468 $\pm$ 0.008& 1.480 $\pm$ 0.005 &0.991\\
        $\text{Seq}_{1}$& 2.509 $\pm$ 0.007& 1.505 $\pm$ 0.004 &0.951\\
        $\text{Seq}_{2}$& 2.502 $\pm$ 0.009& 1.501 $\pm$ 0.005 &0.913\\
        
        \hline
        \end{tabular}
    \caption{Summary of the FH fit results for the coexistence curves. The critical temperature $\text{k}_\text{B}T_\text{c}$, the interaction parameter, $\Delta \varepsilon$, and the coefficient of determination, $R^2$ are listed for each system.}
    \label{tab:SI.FH}
\end{table}
\clearpage


\section{Interfacial density and order-parameter profiles}
To characterize the organization of HP subunits at the dense--dilute interface, we analyzed total monomer number density profiles, chain-end density profiles, the local H/P density contrast, and the orientational order parameter. All profiles are reported as a function of the distance $x$ from the center of the dense phase and averaged over the two interfaces. The position of the interface center, $x_0$, is extracted from the hyperbolic-tangent fit to the density profile (see below and Methods section of the main text).

\subsection{Total monomer number density profiles}
\begin{figure}[htb]
    \centering
    \includegraphics[width=\textwidth]{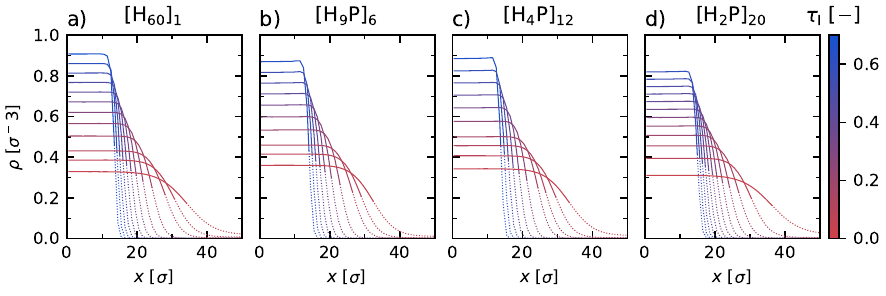}
   \caption{Total monomer number density profile, $\rho(x)$, for the a) $[\text{H}_{60}]_{1}$, b) $[\text{H}_{9}\text{P}]_{6}$, c) $[\text{H}_{4}\text{P}]_{12}$, and d) $[\text{H}_{2}\text{P}]_{20}$ systems as a function of the distance from the center of the dense phase, $x$, averaged over the two interfaces. Individual curves show $\rho(x)$ at various reduced temperatures, $\tau_\text{I}$, as indicated by the color bar. The line style changes from solid in the dense phase to dotted in the dilute phase at the interface center, $x_0$.}
    \label{fig:SI.monomer.number.density.regular}
\end{figure}

\begin{figure}[htb]
    \centering
    \includegraphics[width=\textwidth]{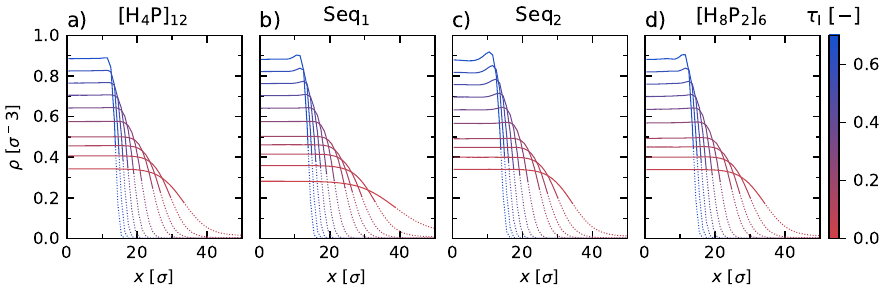}
    \caption{Total monomer number density profile, $\rho(x)$, for the a) $[\text{H}_{4}\text{P}]_{12}$, b) $\text{Seq}_{1}$, c) $\text{Seq}_{2}$, and d) $[\text{H}_{8}\text{P}_{2}]_{6}$ systems as a function of the distance from the center of the dense phase, $x$, averaged over the two interfaces. Individual curves show $\rho(x)$ at various reduced temperatures, $\tau_\text{I}$, as indicated by the color bar. The line style changes from solid in the dense phase to dotted in the dilute phase at the interface center, $x_0$.}
    \label{fig:SI.monomer.number.density.sequence}
\end{figure}

Figure~\ref{fig:SI.monomer.number.density.regular} shows the total monomer number density, $\rho(x)$, for the regular systems of type $\text{H}_n\text{P}$. For these systems, the monomer density changes monotonically across the interface and is well described by a hyperbolic tangent
\begin{equation}
    \rho(x) = \frac{\rho_\text{l} + \rho_\text{v}}{2} + \frac{\rho_\text{l} - \rho_\text{v}}{2} \tanh \left( \frac{x - x_\text{0}}{d}\right),
    \label{eq:tanh}
\end{equation}
where $\rho_\text{l}$ and $\rho_\text{v}$ are the densities in the dense (``liquid'') and dilute (``vapor'') phase, respectively, $x_0$ is the center of the interface, and $d$ the interfacial width. For different sequences at fixed composition, $f_\text{P}=1/5$, $\rho(x)$ qualitatively differs near the interface (Fig.~\ref{fig:SI.monomer.number.density.sequence}), where $\text{Seq}_1$, $\text{Seq}_2$, and $[\text{H}_8\text{P}_2]_{6}$ show a clear increase in monomer number density at the interface relative to their respective bulk values. This feature is most pronounced at low temperatures and indicates the formation of a dense interfacial layer.

\clearpage

\subsection{Composition-normalized H/P contrast}

To characterize deviations in H/P residue fractions close to the interface from the overall composition, we define the local excess of hydrophobic residues,
\begin{equation}
    \Delta \rho (x) = \rho_\text{H}(x)/(1-f_\text{P}) - \rho_\text{P}(x)/f_\text{P},
    \label{eq:SI.delta.rho}
\end{equation}
where $\rho_\text{H}(x)$ and $\rho_\text{P}(x)$ are the monomer number density profiles of H and P residues, respectively. The H and P fractions, $1-f_\text{P}$ and $f_\text{P}$, are used to normalize the densities such that $\Delta \rho = 0$ corresponds to the expected bulk composition. Under this convention, $\Delta \rho < 0$ indicates a surplus of polar residues, whereas $\Delta \rho > 0$ indicates a surplus of hydrophobic residues.

\begin{figure}[htbp]
    \centering
    \includegraphics[width=\textwidth]{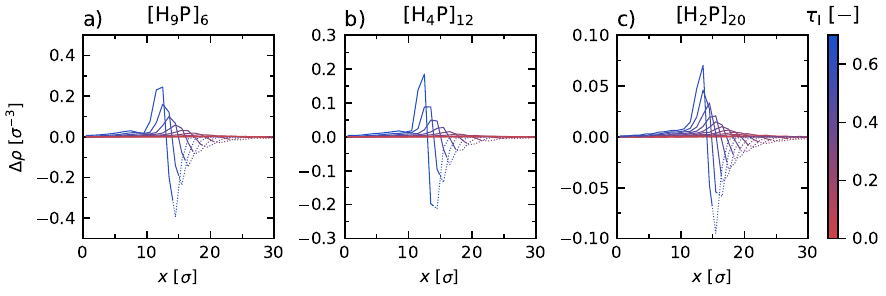}
    \caption{Local excess of hydrophobic and polar residues, $\Delta \rho(x)$, as a function of the distance from the center of the dense phase, $x$, averaged over both interfaces. Data shown for a) $[\text{H}_{9}\text{P}]_{6}$, b) $[\text{H}_{4}\text{P}]_{12}$, and c) $[\text{H}_{2}\text{P}]_{20}$. Individual curves show $\Delta \rho(x)$ at various reduced temperatures, $\tau_\text{I}$, as indicated by the color bar. The line style changes from solid in the dense phase to dotted in the dilute phase at the interface center, $x_0$. Panels show .}
    \label{fig:SI.HP_excess_sequence.regular}
\end{figure}

The corresponding profiles for the regular $\text{H}_n\text{P}$ systems are shown in Figure~\ref{fig:SI.HP_excess_sequence.regular}. These curves exhibit a clear oscillatory H/P pattern near the interface, with $\Delta \rho < 0$ on the dilute side, where polar residues accumulate, and $\Delta \rho > 0$ on the dense side, where hydrophobic residues are enriched. At lower temperatures, a modest excess of polar residues extends further into the interfacial region, before approaching the bulk composition at the center of the dense phase. Among the regular systems, $[\text{H}_9\text{P}]_{6}$ shows the largest interfacial deviation from bulk composition, while the magnitude of the excess decreases with increasing polar fraction. This trend should be interpreted with care, however, because the average polar fraction $f_\text{P}$ is removed from $\Delta\rho(x)$ by construction. Although $[\text{H}_9\text{P}]_{6}$ exhibits a larger relative enrichment of polar residues at the interface, the absolute number of polar residues there remains smaller than in $[\text{H}_2\text{P}]_{20}$, due to its lower overall polar content.

\begin{figure}[htbp]
    \centering
    \includegraphics[width=\textwidth]{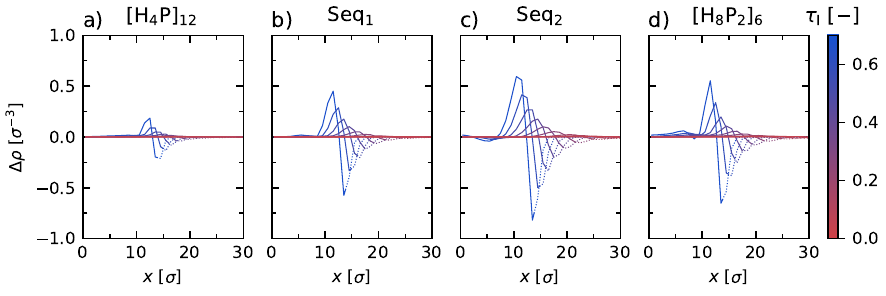}
    \caption{Local excess of hydrophobic and polar residues, $\Delta \rho(x)$, as a function of the distance from the center of the dense phase, $x$, averaged over both interfaces. Data shown for different sequences at fixed composition, $f_\text{P}=1/5$, a) $[\text{H}_{4}\text{P}]_{12}$, b) $\text{Seq}_{1}$, c) $\text{Seq}_2$, and d) $[\text{H}_{8}\text{P}_{2}]_{6}$. Individual curves show $\Delta \rho(x)$ at various reduced temperatures, $\tau_\text{I}$, as indicated by the color bar. The line style changes from solid in the dense phase to dotted in the dilute phase at the interface center, $x_0$. Panels show .}
    \label{fig:SI.HP_excess_sequence}
\end{figure}

The $\Delta \rho(x)$ profiles for the sequence-modulated systems at fixed composition $f_\text{P} = 1/5$ are shown in Fig.~\ref{fig:SI.HP_excess_sequence}. The $[\text{H}_4\text{P}]_{12}$ system exhibits the smallest interfacial H/P contrast, whereas the sequence-modulated systems show substantially stronger interfacial enrichment. In particular, $\text{Seq}_2$ displays long-range oscillations in the H/P pattern, including an additional P-rich region deeper in the dense phase. The magnitude of the P-rich interfacial minimum correlates with the surface tension at fixed composition: systems with more pronounced interfacial deviations from bulk composition tend to exhibit lower surface tension.

\clearpage

\subsection{End-monomer number density profiles}
\begin{figure}[htbp]
    \centering
    \includegraphics[width=\textwidth]{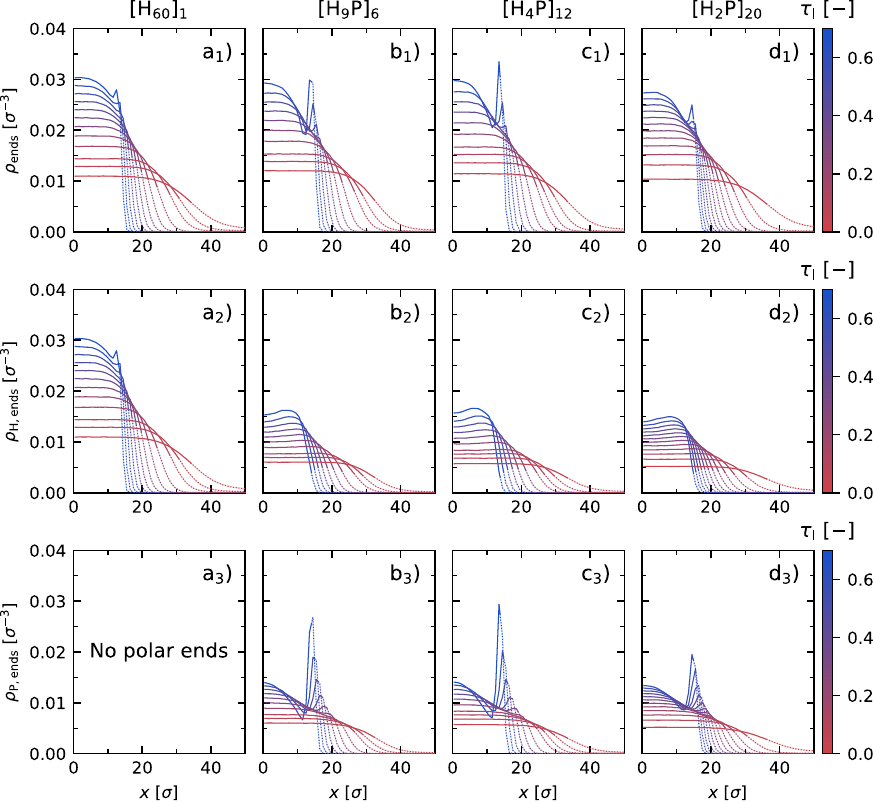}
    \caption{Monomer number density of the end residues (upper panels), hydrophobic end residues (central panels) and polar end residues (lower panels), for regular sequences of type [$\text{H}_n\text{P}$].}
    \label{fig:SI.rho_ends.regular}
\end{figure}

\begin{figure}[htbp]
    \centering
    \includegraphics[width=\textwidth]{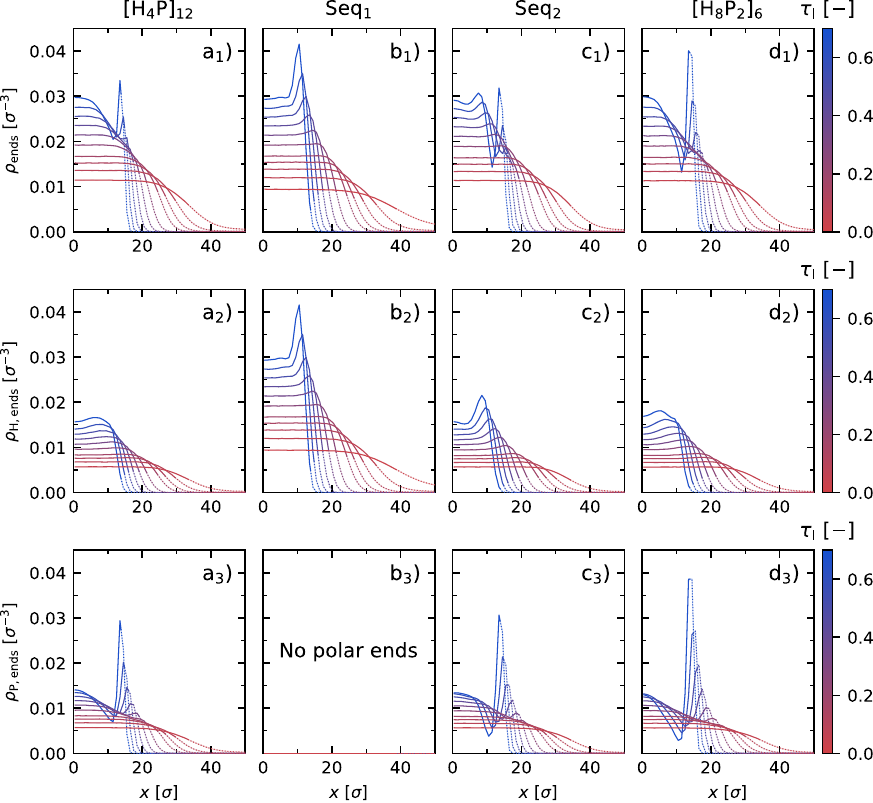}
    \caption{Monomer number density of the end residues (upper panels), hydrophobic end residues (central panels) and polar end residues (lower panels), for different sequences with $f_\text{P}=1/5$.}
    \label{fig:SI.rho_ends.sequence}
\end{figure}

Figures~\ref{fig:SI.rho_ends.regular} and \ref{fig:SI.rho_ends.sequence} show the distribution of end residues within the condensed phase at all investigated (reduced) temperatures.

\clearpage

\subsection{Order Parameter}
To quantify the orientational preference of sequence blocks at the interface, we define a local order parameter based on the vector connecting consecutive H and P residues along the polymer backbone. For each H--P contact along the chain contour, the normalized bond vector is taken to point from H to P and is denoted by $\hat{\mathbf{r}}_{\mathrm{HP}}$. The local interfacial orientation is then measured with respect to the interface normal, $\hat{\mathbf{x}}$, through
\begin{equation}
    \cos\theta(x) = \hat{\mathbf{r}}_{\mathrm{HP}} \cdot \hat{\mathbf{x}}.
    \label{eq:SI.cos_theta}
\end{equation}
By construction, $\cos\theta(x) > 0$ indicates a preferential orientation from H toward P along the interface normal, whereas $\cos\theta(x) < 0$ indicates the opposite orientation.

\begin{figure}[htb]
    \centering
    \includegraphics[width=\textwidth]{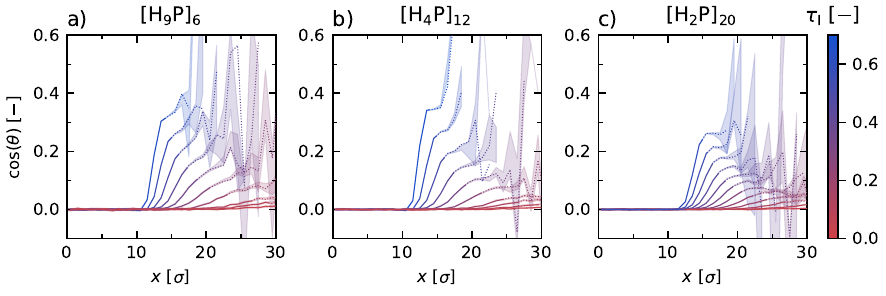}
    \caption{Order parameter, $\cos \theta(x)$, for regular $\text{H}_{n}\text{P}$ systems as a function of the distance from the center of the dense phase, $x$, averaged over both interfaces. Data shown for a) $[\text{H}_{9}\text{P}]_{6}$, b) $[\text{H}_{4}\text{P}]_{12}$, and c) $[\text{H}_{2}\text{P}]_{20}$. Individual curves show $\cos \theta(x)$ at various reduced temperatures, $\tau_\text{I}$, as indicated by the color bar. The line style changes from solid in the dense phase to dotted in the dilute phase at the interface center, $x_0$, and filled regions mark the standard error of the mean for each curve.}
    \label{fig:SI.cos_theta_regular}
\end{figure}

The corresponding profiles for the regular $\text{H}_n\text{P}$ systems reveal that close to the interface, the HP subunits orient with respect to the interface such that the polar residues point toward the dilute phase and the hydrophobic residues toward the dense phase (Figure~\ref{fig:SI.cos_theta_regular}). This orientational preference is strongest on the dilute-side of the interface, but remains substantial on the dense side as well.

The sequence-modulated systems at fixed composition exhibit a qualitatively different behavior. In particular, $\cos\theta(x) < 0$ in the dense phase, indicating that the HP subunits align the polar residues away from the interface, indicative of a second polar layer, as has been seen from the $\Delta \rho$ analysis in Figure~\ref{fig:SI.HP_excess_sequence}. Away from this region, however, their behavior is similar to that of the regular sequences: the HP subunits orient at the interface to expose the polar residues towards the dilute phase while anchoring the hydrophobic residues in the dense phase.

\begin{figure}[htb]
    \centering
    \includegraphics[width=\textwidth]{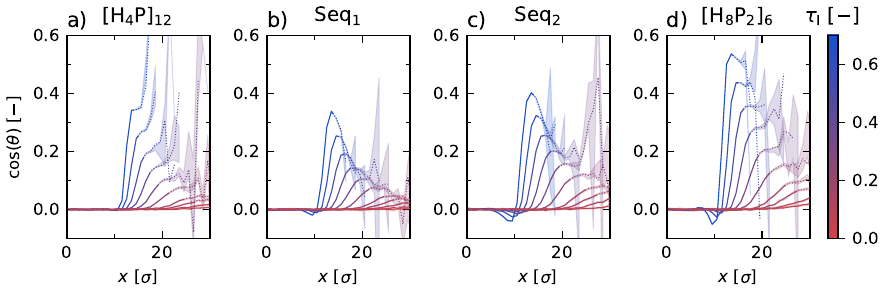}
    \caption{Order parameter, $\cos \theta(x)$, as a function of the distance from the center of the dense phase, $x$, averaged over both interfaces. Data shown for different sequences at fixed composition, $f_\text{P}=1/5$, a) $[\text{H}_{4}\text{P}]_{12}$, b) $\text{Seq}_{1}$, c) $\text{Seq}_2$, and d) $[\text{H}_{8}\text{P}_{2}]_{6}$. Individual curves show $\cos \theta(x)$ at various reduced temperatures, $\tau_\text{I}$, as indicated by the color bar. The line style changes from solid in the dense phase to dotted in the dilute phase at the interface center, $x_0$, and filled regions mark the standard error of the mean for each curve.}
    \label{fig:SI.cos_theta_sequence}
\end{figure}

\clearpage
\section{Critical behavior and surface tension}
The parameters obtained from the Ising-scaling analysis and the associated coefficients of determination are summarized in Table~\ref{tab:SI.surface_tension}. For each system, the critical temperature, $T_\text{c,I}$ was obtained by fitting the density difference $\Delta \rho$ at the four highest temperatures to the scaling relation
\begin{equation}
    \Delta\rho = \rho_\text{l} - \rho_\text{v}
    = A\tau^\beta,
\end{equation}
where $\beta \approx 0.326$ is the critical exponent of the three-dimensional Ising universality class. The critical amplitude of the surface tension, $\gamma_0$, was obtained by fitting $\gamma$ to Eq.~\eqref{eq:gamma} over the range $\tau_\text{I} \leq 0.5$, where the systems follow the asymptotic power-law behavior. The difference in surface tension relative to the homopolymer
\begin{equation}
    \Delta \gamma(\tau_\text{I}) = \gamma_{\text{H}_{60}}(\tau_\text{I}) - \gamma_{\text{H}_n\text{P}}(\tau_\text{I}).
\end{equation}
was fitted to a quadratic function in $\tau_\text{I}$ over the range $\tau_\text{I} \geq 0.3$.

\begin{table}[htbp]
    \centering
    \setlength{\tabcolsep}{12pt}
    \renewcommand{\arraystretch}{1.05}
    \begin{tabular}{lcccccc}
        \hline
        System&  $T_\text{c,I}$& $R^2$ & $\gamma_0$ [$\varepsilon/\sigma^2$] & $R^2$ & $\Delta \gamma_0$ [$\varepsilon/\sigma^2$] & $R^2$\\
        \hline
        $[\text{H}_{60}]_1$& 3.112 $\pm$ 0.002& 1.000& 1.593 $\pm$ 0.007& 0.996& - & -\\
        $[\text{H}_{9}\text{P}]_{6}$ &  2.743 $\pm$ 0.003& 0.999& 1.298 $\pm$ 0.004& 0.999& 0.599 $\pm$ 0.004& 0.994\\
        $[\text{H}_{4}\text{P}]_{12}$ & 2.408 $\pm$ 0.002& 0.999& 1.118 $\pm$ 0.003& 0.999& 0.915 $\pm$ 0.005& 0.993\\
        $[\text{H}_{2}\text{P}]_{20}$ & 1.966 $\pm$ 0.001& 0.999&  0.874 $\pm$ 0.002& 0.999& 1.312 $\pm$ 0.011& 0.974\\
        $[\text{H}_{8}\text{P}_{2}]_{6}$ & 2.411 $\pm$ 0.003& 0.998& 0.904 $\pm$ 0.004& 0.997& 1.464 $\pm$ 0.017& 0.993\\
        $\text{Seq}_{1}$ & 2.456 $\pm$ 0.001& 1.000& 1.031 $\pm$ 0.003& 0.998& 1.196 $\pm$ 0.014& 0.993\\
        $\text{Seq}_{2}$ & 2.417 $\pm$ 0.003& 0.998& 0.854 $\pm$ 0.006& 0.992& 1.624 $\pm$ 0.019& 0.994\\
        \hline
        \end{tabular}
    \caption{Critical temperature, $T_\text{c,I}$, critical amplitude of the surface tension, $\gamma_0$, and the prefactor of the quadratic fit to the surface tension difference between homo- and heteropolymer systems, $\Delta \gamma_0$. All data include their respective coefficient of determination, $R^2$.}
    \label{tab:SI.surface_tension}
\end{table}
\clearpage

\section{Additional sequences}

\begin{figure}[htbp]
    \centering
    \includegraphics[width=\textwidth]{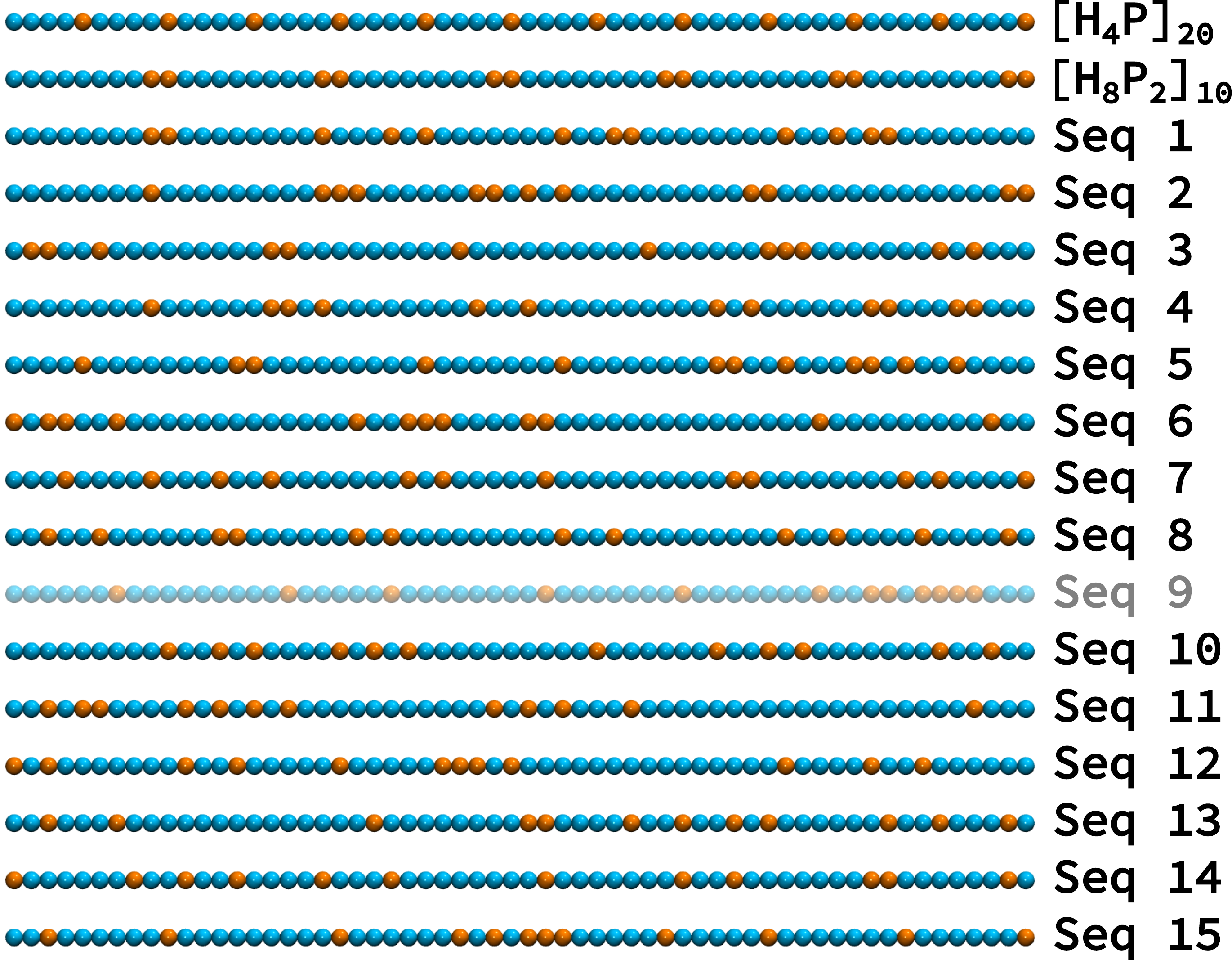}
    \caption{Heteropolymers with equal composition $f_\text{P}=1/5$ but different sequences. Hydrophobic and polar residues are shown in blue and orange, respectively. $\text{Seq}_9$ was excluded from our analysis, as discussed in the text.}
    \label{fig:SI.sequences}
\end{figure}

To further investigate how sequence patterning affects surface tension, we simulated 13 additional disordered sequences at fixed composition $f_\text{P}=1/5$ (Fig.~\ref{fig:SI.sequences}). Constructing a full phase diagram for each sequence is computationally expensive. However, the four sequence variants discussed in the main text ($[\text{H}_4\text{P}]_{12}$, $[\text{H}_8\text{P}_2]_{6}$, $\text{Seq}_1$, and $\text{Seq}_2$) exhibit nearly identical phase behavior. We therefore simulated and analyzed the additional sequences at a single temperature, $k_\text{B}T=0.8\varepsilon$.

\begin{figure}[htbp]
    \centering
    \includegraphics[width=\textwidth]{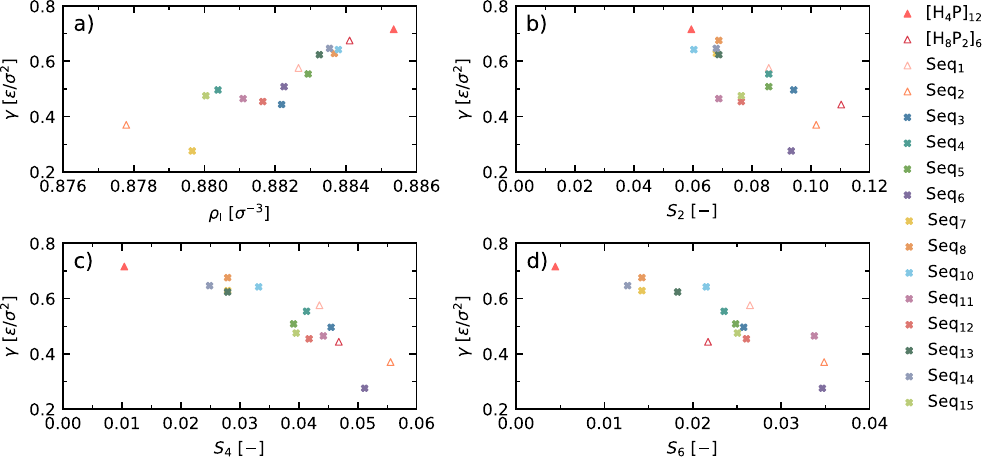}
    \caption{Surface tension $\gamma$ as a function of  (a) dense phase density, $\rho_\text{l}$, and composition heterogeneity $S_\omega$ for window sizes (b) $\omega=2$, (c) $\omega=4$, and (d) $\omega=6$.}
    \label{fig:SI.sequence.dependence}
\end{figure}

The dense-phase densities $\rho_\text{l}$ of the additional sequences lie between those of $\text{Seq}_1$ and $[\text{H}_4\text{P}]_{12}$ (Fig.~\ref{fig:SI.sequence.dependence}a), with the exception of $\text{Seq}_9$. For this sequence, polar residues localize strongly at the interface, and the monomer number density oscillates deep into the dense phase; therefore, this sequence was excluded from the subsequent analysis. For the remaining sequences, the measured surface tensions range from $\gamma=0.276\pm0.003$ for $\text{Seq}_6$ to $\gamma=0.716\pm0.001$ for $[\text{H}_4\text{P}]_{12}$.

We characterize the blockiness of polar residues at fixed composition through the parameter
\begin{equation}
    S_\omega = \frac{1}{M-\omega+1} \sum_{i=1}^{M-\omega} \left( f_{\text{P},i}^\omega - f_\text{P}\right)^2,
\end{equation}
which measures how strongly the local P fraction varies along a sub-sequence of $\omega$ consecutive residues. For each window of length $\omega$, the local P fraction $f_{\text{P},i}^\omega$ is compared with the overall P fraction $f_\text{P}$. The squared deviations $\left( f_{\text{P},i}^\omega - f_\text{P}\right)^2$ are then averaged over all windows. Thus, a larger $S_\omega$ indicates more local clustering of P residues.
Setting $\omega=2$, $S_2$ probes the arrangement of immediate P-P neighbors along the chain, while larger $\omega$ values probe clustering in longer sub-sequences. These larger windows can capture P-rich regions even when P beads are separated by one or more H beads. Figure~\ref{fig:SI.sequence.dependence}b-d show $S_2$, $S_4$, and $S_6$ for all sequences except $\text{Seq}_9$, indicating a tendency towards lower surface tension with increasing composition heterogeneity. Note, that the magnitude of $S_\omega$ depends on the window size, so its absolute values should not be compared across panels.

\clearpage

\section{System specific parameters}

The following tables report the system-specific values for the seven systems studied in the main text. For each system, we list the simulation temperature $T$, reduced temperature $\tau_\text{I}$, dense- and dilute-phase densities $\rho_\text{l}$ and $\rho_\text{v}$, respectively, interface center $x_0$, interfacial width $d$, and surface tension $\gamma$. The reduced temperature was calculated using the critical temperature $T_\text{c}$ obtained from the Ising fit. 

Uncertainties in $\rho_\text{l}$, $\rho_\text{v}$, $x_0$, and $d$ were estimated by block averaging. Each trajectory was divided into five blocks containing 2000 snapshots, corresponding to a block length of $400{,}000\,\tau_\text{MD}$, and the standard error across the blocks is reported in the tables. Uncertainties shown as zero are smaller than the displayed numerical precision.

\begin{table}[htbp]
    \centering
    \setlength{\tabcolsep}{5pt}
    \renewcommand{\arraystretch}{1.05}
    \begin{tabular}{ccccccl}
        \hline
         $T$ [$\varepsilon / \text{k}_\text{B}$]& $\tau_{\text{I}}$& $\rho_{\text{l}}$ [$\sigma^{-3}$]& $\rho_{\text{v}}$ [$\sigma^{-3}$]& $x_0$ [$\sigma$]& $d$ [$\sigma$]&$\gamma$ [$\varepsilon / \sigma^2$]\\
        \hline
         1.0 & 0.679 $\pm$ 0.000& 0.91 $\pm$ 0.00& -& 13.51 $\pm$ 0.00& 0.87 $\pm$ 0.00&1.162 $\pm$ 0.001\\
         1.2 & 0.614 $\pm$ 0.000& 0.86 $\pm$ 0.00& -& 14.25 $\pm$ 0.00& 1.03 $\pm$ 0.00&0.976 $\pm$ 0.001\\
         1.4 & 0.550 $\pm$ 0.000& 0.81 $\pm$ 0.00& -& 15.06 $\pm$ 0.00& 1.19 $\pm$ 0.00&0.814 $\pm$ 0.001\\
         1.6 & 0.486 $\pm$ 0.000& 0.77 $\pm$ 0.00& -& 15.97 $\pm$ 0.00& 1.37 $\pm$ 0.00&0.668 $\pm$ 0.001\\
         1.8 & 0.422 $\pm$ 0.000& 0.72 $\pm$ 0.00& -& 17.01 $\pm$ 0.00& 1.58 $\pm$ 0.00&0.538 $\pm$ 0.002\\
         2.0 & 0.357 $\pm$ 0.000& 0.67 $\pm$ 0.00& -& 18.24 $\pm$ 0.00& 1.83 $\pm$ 0.00&0.422 $\pm$ 0.002\\
         2.2 & 0.293 $\pm$ 0.000& 0.62 $\pm$ 0.00& -& 19.74 $\pm$ 0.00& 2.16 $\pm$ 0.00&0.322 $\pm$ 0.001\\
         2.4 & 0.229 $\pm$ 0.000& 0.57 $\pm$ 0.00& -& 21.66 $\pm$ 0.00& 2.60 $\pm$ 0.00&0.232 $\pm$ 0.001\\
         2.6 & 0.164 $\pm$ 0.000& 0.50 $\pm$ 0.00& 0.0004 $\pm$ 0.001& 24.24 $\pm$ 0.01& 3.23 $\pm$ 0.01&0.152 $\pm$ 0.003\\
         2.8 & 0.100 $\pm$ 0.000& 0.43 $\pm$ 0.00& 0.0033 $\pm$ 0.003& 27.97 $\pm$ 0.03& 4.36 $\pm$ 0.01&0.081 $\pm$ 0.002\\
         2.9 & 0.068 $\pm$ 0.000& 0.39 $\pm$ 0.00& 0.0072 $\pm$ 0.002& 30.65 $\pm$ 0.03& 5.43 $\pm$ 0.01&0.053 $\pm$ 0.002\\
         3.0 & 0.036 $\pm$ 0.000& 0.33 $\pm$ 0.00& 0.0196 $\pm$ 0.005& 34.10 $\pm$ 0.06& 7.60 $\pm$ 0.09&0.025 $\pm$ 0.002\\
        \hline
    \end{tabular}
    \caption{System-specific parameters for the $[\text{H}_{60}]_1$ system at each temperature. Uncertainties reported as zero indicate that the statistical uncertainty is smaller than the displayed numerical precision.}
    \label{tab:SI.coexistance.data.H60}
\end{table}

\begin{table}[htbp]
    \centering
    \setlength{\tabcolsep}{5pt}
    \renewcommand{\arraystretch}{1.05}
    \begin{tabular}{ccccccl}
        \hline
         $T$ [$\varepsilon / \text{k}_\text{B}$]& $\tau_{\text{I}}$& $\rho_{\text{l}}$ [$\sigma^{-3}$]& $\rho_{\text{v}}$ [$\sigma^{-3}$]& $x_0$ [$\sigma$]& $d$ [$\sigma$]&$\gamma$ [$\varepsilon / \sigma^2$]\\
        \hline
         1.0 & 0.635 $\pm$ 0.000& 0.87 $\pm$ 0.00& -& 14.06 $\pm$ 0.00& 1.01 $\pm$ 0.00&0.788 $\pm $ 0.002\\
         1.2 & 0.563 $\pm$ 0.000& 0.82 $\pm$ 0.00& -& 14.97 $\pm$ 0.00& 1.17 $\pm$ 0.00&0.657 $\pm $ 0.002\\
         1.4 & 0.490 $\pm$ 0.001& 0.76 $\pm$ 0.00& -& 16.00 $\pm$ 0.00& 1.36 $\pm$ 0.00&0.538 $\pm $ 0.002\\
         1.6 & 0.417 $\pm$ 0.001& 0.71 $\pm$ 0.00& -& 17.20 $\pm$ 0.00& 1.60 $\pm$ 0.00&0.431 $\pm $ 0.001\\
         1.8 & 0.344 $\pm$ 0.001& 0.66 $\pm$ 0.00& -& 18.64 $\pm$ 0.00& 1.89 $\pm$ 0.00&0.333 $\pm $ 0.002\\
         2.0 & 0.271 $\pm$ 0.001& 0.60 $\pm$ 0.00& -& 20.46 $\pm$ 0.00& 2.28 $\pm$ 0.00&0.239 $\pm $ 0.001\\
         2.2 & 0.198 $\pm$ 0.001& 0.53 $\pm$ 0.00& -& 22.92 $\pm$ 0.00& 2.84 $\pm$ 0.00&0.162 $\pm $ 0.002\\
         2.4 & 0.125 $\pm$ 0.001& 0.46 $\pm$ 0.00& 0.0012 $\pm$ 0.0002& 26.49 $\pm$ 0.02& 3.78 $\pm$ 0.01&0.093 $\pm $ 0.002\\
         2.5 & 0.089 $\pm$ 0.001& 0.41 $\pm$ 0.00& 0.0039 $\pm$ 0.0002& 28.95 $\pm$ 0.02& 4.58 $\pm$ 0.01&0.060 $\pm $ 0.003\\
         2.6 & 0.052 $\pm$ 0.001& 0.36 $\pm$ 0.00& 0.0103 $\pm$ 0.0004& 32.38 $\pm$ 0.06& 6.03 $\pm$ 0.05&0.033 $\pm $ 0.001\\
        \hline
    \end{tabular}
    \caption{System-specific parameters for the $[\text{H}_{9}\text{P}]_{6}$ system at each temperature. Uncertainties reported as zero indicate that the statistical uncertainty is smaller than the displayed numerical precision.}
    \label{tab:SI.coexistance.data.H9P}
\end{table}

\begin{table}[htbp]
    \centering
    \setlength{\tabcolsep}{5pt}
    \renewcommand{\arraystretch}{1.05}
    \begin{tabular}{ccccccl}
        \hline
         $T$ [$\varepsilon / \text{k}_\text{B}$]& $\tau_{\text{I}}$& $\rho_{\text{l}}$ [$\sigma^{-3}$]& $\rho_{\text{v}}$ [$\sigma^{-3}$]& $x_0$ [$\sigma$]& $d$ [$\sigma$]&$\gamma$ [$\varepsilon / \sigma^2$]\\
        \hline
         0.8& 0.668 $\pm$ 0.000& 0.89 $\pm$ 0.00& -& 13.81 $\pm$ 0.00& 0.93 $\pm$ 0.00&0.716 $\pm$ 0.01\\
         1.0& 0.585 $\pm$ 0.000& 0.83 $\pm$ 0.00& -& 14.83 $\pm$ 0.00& 1.11 $\pm$ 0.00&0.597 $\pm$ 0.01\\
         1.2& 0.502 $\pm$ 0.000& 0.77 $\pm$ 0.00& -& 15.98 $\pm$ 0.00& 1.33 $\pm$ 0.00&0.476 $\pm$ 0.02\\
         1.4& 0.419 $\pm$ 0.000& 0.71 $\pm$ 0.00& -& 17.35 $\pm$ 0.00& 1.60 $\pm$ 0.00&0.374 $\pm$ 0.02\\
         1.6& 0.336 $\pm$ 0.001& 0.64 $\pm$ 0.00& -& 19.04 $\pm$ 0.00& 1.93 $\pm$ 0.00&0.276 $\pm$ 0.01\\
         1.8& 0.252 $\pm$ 0.001& 0.58 $\pm$ 0.00& -& 21.25 $\pm$ 0.00& 2.41 $\pm$ 0.00&0.189 $\pm$ 0.01\\
         2.0& 0.169 $\pm$ 0.001& 0.50 $\pm$ 0.00& 0.0003 $\pm$ 0.0001& 24.43 $\pm$ 0.01& 3.16 $\pm$ 0.01&0.116 $\pm$ 0.02\\
         2.1& 0.128 $\pm$ 0.001& 0.46 $\pm$ 0.00& 0.0011 $\pm$ 0.0002& 26.65 $\pm$ 0.02& 3.74 $\pm$ 0.01&0.082 $\pm$ 0.02\\
         2.2& 0.086 $\pm$ 0.001& 0.41 $\pm$ 0.00& 0.0038 $\pm$ 0.0004& 29.52 $\pm$ 0.06& 4.71 $\pm$ 0.02&0.052 $\pm$ 0.01\\
         2.3& 0.045 $\pm$ 0.001& 0.34 $\pm$ 0.00& 0.0152 $\pm$ 0.0011& 33.32 $\pm$ 0.14& 6.68 $\pm$ 0.04&0.026 $\pm$ 0.01\\
        \hline
    \end{tabular}
    \caption{System-specific parameters for the $[\text{H}_{4}\text{P}]_{12}$ system at each temperature. Uncertainties reported as zero indicate that the statistical uncertainty is smaller than the displayed numerical precision.}
    \label{tab:SI.coexistance.data.H4P}
\end{table}

\begin{table}[htbp]
    \centering
    \setlength{\tabcolsep}{5pt}
    \renewcommand{\arraystretch}{1.05}
    \begin{tabular}{ccccccl}
        \hline
         $T$ [$\varepsilon / \text{k}_\text{B}$]& $\tau_{\text{I}}$& $\rho_{\text{l}}$ [$\sigma^{-3}$]& $\rho_{\text{v}}$ [$\sigma^{-3}$]& $x_0$ [$\sigma$]& $d$ [$\sigma$]&$\gamma$ [$\varepsilon / \sigma^2$]\\
        \hline
         0.8& 0.668 $\pm$ 0.000& 0.88 $\pm$ 0.00& -& 13.79 $\pm$ 0.00& 1.04 $\pm$ 0.00&0.444 $\pm$ 0.001\\
         1.0& 0.585 $\pm$ 0.000& 0.82 $\pm$ 0.00& -& 14.83 $\pm$ 0.00& 1.19 $\pm$ 0.00&0.414 $\pm$ 0.001\\
         1.2& 0.502 $\pm$ 0.001& 0.76 $\pm$ 0.00& -& 16.02 $\pm$ 0.00& 1.38 $\pm$ 0.00&0.363 $\pm$ 0.002\\
         1.4& 0.419 $\pm$ 0.001& 0.70 $\pm$ 0.00& -& 17.43 $\pm$ 0.00& 1.62 $\pm$ 0.00&0.296 $\pm$ 0.001\\
         1.6& 0.336 $\pm$ 0.001& 0.64 $\pm$ 0.00& -& 19.17 $\pm$ 0.00& 1.94 $\pm$ 0.00&0.229 $\pm$ 0.002\\
         1.8& 0.253 $\pm$ 0.001& 0.57 $\pm$ 0.00& -& 21.44 $\pm$ 0.00& 2.41 $\pm$ 0.00&0.165 $\pm$ 0.001\\
         2.0& 0.171 $\pm$ 0.001& 0.49 $\pm$ 0.00& 0.0003 $\pm$ 0.002& 24.71 $\pm$ 0.01& 3.12 $\pm$ 0.01&0.100 $\pm$ 0.001\\
         2.1& 0.129 $\pm$ 0.001& 0.45 $\pm$ 0.00& 0.0008 $\pm$ 0.001& 27.00 $\pm$ 0.02& 3.69 $\pm$ 0.02&0.076 $\pm$ 0.001\\
         2.2& 0.088 $\pm$ 0.001& 0.40 $\pm$ 0.00& 0.0031 $\pm$ 0.002& 30.04 $\pm$ 0.03& 4.63 $\pm$ 0.01&0.046 $\pm$ 0.002\\
         2.3& 0.046 $\pm$ 0.001& 0.34 $\pm$ 0.00& 0.0125$\pm$ 0.014& 34.12 $\pm$ 0.20& 6.49 $\pm$ 0.06&0.025 $\pm$ 0.001\\
        \hline
    \end{tabular}
    \caption{System-specific parameters for the $[\text{H}_{2}\text{P}]_{20}$ system at each temperature. Uncertainties reported as zero indicate that the statistical uncertainty is smaller than the displayed numerical precision.}
    \label{tab:SI.coexistance.data.H8P2}
\end{table}

\begin{table}[htbp]
    \centering
    \setlength{\tabcolsep}{5pt}
    \renewcommand{\arraystretch}{1.05}
    \begin{tabular}{ccccccl}
        \hline
         $T$ [$\varepsilon / \text{k}_\text{B}$]& $\tau_{\text{I}}$& $\rho_{\text{l}}$ [$\sigma^{-3}$]& $\rho_{\text{v}}$ [$\sigma^{-3}$]& $x_0$ [$\sigma$]& $d$ [$\sigma$]&$\gamma$ [$\varepsilon / \sigma^2$]\\
        \hline
         0.8& 0.674 $\pm$ 0.000& 0.88 $\pm$ 0.00& -& 13.81 $\pm$ 0.00& 0.99 $\pm$ 0.00&0.576 $\pm$ 0.002\\
         1.0& 0.593 $\pm$ 0.000& 0.82 $\pm$ 0.00& -& 14.83 $\pm$ 0.00& 1.14 $\pm$ 0.00&0.512 $\pm$ 0.002\\
         1.2& 0.512 $\pm$ 0.000& 0.76 $\pm$ 0.00& -& 16.00 $\pm$ 0.00& 1.33 $\pm$ 0.00&0.434 $\pm$ 0.001\\
         1.4& 0.430 $\pm$ 0.000& 0.70 $\pm$ 0.00& -& 17.38 $\pm$ 0.00& 1.57 $\pm$ 0.00&0.351 $\pm$ 0.002\\
         1.6& 0.349 $\pm$ 0.000& 0.64 $\pm$ 0.00& -& 19.06 $\pm$ 0.00& 1.89 $\pm$ 0.00&0.273 $\pm$ 0.002\\
         1.8& 0.267 $\pm$ 0.000& 0.58 $\pm$ 0.00& -& 21.26 $\pm$ 0.00& 2.34 $\pm$ 0.00&0.199 $\pm$ 0.001\\
 2.0& 0.186 $\pm$ 0.000& 0.50 $\pm$ 0.00& 0.0001 $\pm$ 0.0000& 24.34 $\pm$ 0.00& 3.04 $\pm$ 0.01&0.125 $\pm$ 0.001\\
         2.1& 0.145 $\pm$ 0.000& 0.46 $\pm$ 0.00& 0.0008 $\pm$ 0.0002& 26.43 $\pm$ 0.01& 3.59 $\pm$ 0.00&0.095 $\pm$ 0.001\\
         2.2& 0.104 $\pm$ 0.000& 0.42 $\pm$ 0.00& 0.0025 $\pm$ 0.0003& 29.11 $\pm$ 0.03& 4.40 $\pm$ 0.02&0.067 $\pm$ 0.002\\
         2.3& 0.064 $\pm$ 0.000& 0.36 $\pm$ 0.00& 0.0081 $\pm$ 0.0005& 32.87 $\pm$ 0.04& 5.84 $\pm$ 0.03&0.040 $\pm$ 0.002\\
         2.4& 0.023 $\pm$ 0.000& 0.28 $\pm$ 0.00& 0.0273 $\pm$ 0.0008& 38.62 $\pm$ 0.16& 9.68 $\pm$ 0.14&0.011 $\pm$ 0.002\\
         \hline
    \end{tabular}
    \caption{System-specific parameters for the $\text{Seq}_{1}$ system at each temperature. Uncertainties reported as zero indicate that the statistical uncertainty is smaller than the displayed numerical precision.}
    \label{tab:SI.coexistance.data.Seq1}
\end{table}

\begin{table}[htbp]
    \centering
    \setlength{\tabcolsep}{5pt}
    \renewcommand{\arraystretch}{1.05}
    \begin{tabular}{ccccccl}
        \hline
         $T$ [$\varepsilon / \text{k}_\text{B}$]& $\tau_{\text{I}}$& $\rho_{\text{l}}$ [$\sigma^{-3}$]& $\rho_{\text{v}}$ [$\sigma^{-3}$]& $x_0$ [$\sigma$]& $d$ [$\sigma$]&$\gamma$ [$\varepsilon / \sigma^2$]\\
        \hline
         0.8& 0.669 $\pm$ 0.000& 0.88 $\pm$ 0.00& -& 13.81 $\pm$ 0.00& 1.10 $\pm$ 0.00&0.370 $\pm$ 0.001\\
         1.0& 0.586 $\pm$ 0.001& 0.82 $\pm$ 0.00& -& 14.86 $\pm$ 0.00& 1.23 $\pm$ 0.00&0.358 $\pm$ 0.001\\
         1.2& 0.504 $\pm$ 0.001& 0.76 $\pm$ 0.00& -& 16.05 $\pm$ 0.00& 1.40 $\pm$ 0.00&0.322 $\pm$ 0.002\\
         1.4& 0.421 $\pm$ 0.001& 0.70 $\pm$ 0.00& -& 17.48 $\pm$ 0.00& 1.62 $\pm$ 0.00&0.277 $\pm$ 0.001\\
         1.6& 0.338 $\pm$ 0.001& 0.63 $\pm$ 0.00& -& 19.24 $\pm$ 0.00& 1.93 $\pm$ 0.00&0.219 $\pm$ 0.002\\
         1.8& 0.255 $\pm$ 0.001& 0.57 $\pm$ 0.00& -& 21.55 $\pm$ 0.00& 2.38 $\pm$ 0.00&0.160 $\pm$ 0.001\\
         2.0& 0.173 $\pm$ 0.001& 0.49 $\pm$ 0.00& 0.0002 $\pm$ 0.0000& 24.85 $\pm$ 0.01& 3.09 $\pm$ 0.01&0.101 $\pm$ 0.001\\
         2.1& 0.131 $\pm$ 0.001& 0.45 $\pm$ 0.00& 0.0008 $\pm$ 0.0001& 27.14 $\pm$ 0.02& 3.65 $\pm$ 0.01&0.073 $\pm$ 0.001\\
         2.2& 0.090 $\pm$ 0.001& 0.40 $\pm$ 0.00& 0.0033 $\pm$ 0.0004& 30.10 $\pm$ 0.04& 4.59 $\pm$ 0.02&0.050 $\pm$ 0.002\\
         2.3& 0.049 $\pm$ 0.001& 0.34 $\pm$ 0.00& 0.0105 $\pm$ 0.0008& 34.25 $\pm$ 0.11& 6.27 $\pm$ 0.05&0.025 $\pm$ 0.001\\
        \hline
    \end{tabular}
    \caption{System-specific parameters for the $\text{Seq}_{2}$ system at each temperature. Uncertainties reported as zero indicate that the statistical uncertainty is smaller than the displayed numerical precision.}
    \label{tab:SI.coexistance.data.Seq2}
\end{table}

\begin{table}[htbp]
\footnotesize
    \centering
    \setlength{\tabcolsep}{5pt}
    \renewcommand{\arraystretch}{1.05}
    \begin{tabular}{ccccccl}
        \hline
         $T$ [$\varepsilon / \text{k}_\text{B}$]& $\tau_{\text{I}}$& $\rho_{\text{l}}$ [$\sigma^{-3}$]& $\rho_{\text{v}}$ [$\sigma^{-3}$]& $x_0$ [$\sigma$]& $d$ [$\sigma$]&$\gamma$ [$\varepsilon / \sigma^2$]\\
        \hline
         0.8& 0.593 $\pm$ 0.000& 0.82 $\pm$ 0.00& -& 14.92 $\pm$ 0.00& 1.10 $\pm$ 0.00&0.474 $\pm$ 0.001\\
         0.9& 0.542 $\pm$ 0.000& 0.78 $\pm$ 0.00& -& 15.61 $\pm$ 0.00& 1.22 $\pm$ 0.00&0.419 $\pm$ 0.001\\
         1.0& 0.491 $\pm$ 0.000& 0.75 $\pm$ 0.00& -& 16.37 $\pm$ 0.00& 1.37 $\pm$ 0.00&0.364 $\pm$ 0.001\\
         1.1& 0.440 $\pm$ 0.000& 0.71 $\pm$ 0.00& -& 17.21 $\pm$ 0.00& 1.53 $\pm$ 0.00&0.313 $\pm$ 0.001\\
         1.2& 0.390 $\pm$ 0.000& 0.67 $\pm$ 0.00& -& 18.16 $\pm$ 0.00& 1.71 $\pm$ 0.00&0.265 $\pm$ 0.001\\
         1.3& 0.339 $\pm$ 0.000& 0.64 $\pm$ 0.00& -& 19.26 $\pm$ 0.00& 1.94 $\pm$ 0.00&0.216 $\pm$ 0.001\\
         1.4& 0.288 $\pm$ 0.000& 0.60 $\pm$ 0.00& -& 20.55 $\pm$ 0.00& 2.21 $\pm$ 0.00&0.179 $\pm$ 0.001\\
         1.5& 0.237 $\pm$ 0.000& 0.55 $\pm$ 0.00& -& 22.12 $\pm$ 0.00& 2.56 $\pm$ 0.00&0.137 $\pm$ 0.002\\
         1.6& 0.186 $\pm$ 0.001& 0.51 $\pm$ 0.00& 0.0002 $\pm$ 0.0001& 24.11 $\pm$ 0.01& 3.02 $\pm$ 0.01&0.104 $\pm$ 0.002\\
         1.7& 0.135 $\pm$ 0.001& 0.46 $\pm$ 0.00& 0.0008 $\pm$ 0.0001& 26.74 $\pm$ 0.02& 3.68 $\pm$ 0.01&0.067 $\pm$ 0.002\\
         1.8& 0.084 $\pm$ 0.001& 0.39 $\pm$ 0.00& 0.0045 $\pm$ 0.0004& 30.34 $\pm$ 0.04& 4.91 $\pm$ 0.02&0.041 $\pm$ 0.002\\
         1.9& 0.033 $\pm$ 0.001& 0.31 $\pm$ 0.00& 0.0187 $\pm$ 0.0006& 36.20 $\pm$ 0.09& 8.06 $\pm$ 0.14&0.016 $\pm$ 0.001\\
        \hline
    \end{tabular}
    \caption{System-specific parameters for the $[\text{H}_{8}\text{P}_{2}]_{6}$ system at each temperature. Uncertainties reported as zero indicate that the statistical uncertainty is smaller than the displayed numerical precision.}
    \label{tab:SI.coexistance.data.H2P}
\end{table}

\clearpage
\bibliography{references}